\documentclass[preprint,authoryear,onecolumn,12pt]{elsarticle}
\usepackage{amssymb}  
\usepackage{amsfonts}
\usepackage{times}   
\usepackage{graphicx}  
\usepackage{float}
\usepackage{indentfirst} 
\usepackage{amsmath}
\usepackage{epstopdf}
\usepackage{color}
\usepackage{textcomp}
\usepackage{subfigure}
\usepackage[all]{xy}
\usepackage{longtable}
\usepackage{amssymb}
\usepackage{animate}

\begin{document}

\begin{frontmatter}
\title{Selfish punishment with avoiding mechanism can alleviate both first-order and second-order social dilemma\tnoteref{label1}}
%\tnotetext[label1]{}
\author[label1,label2]{Pengbi Cui}
\author[label1,label2]{Zhi-Xi Wu\corref{cor1}}
\ead{wuzhx@lzu.edu.cn}
\cortext[cor1]{Corresponding author at: Institute of Computational Physics and Complex
Systems, Lanzhou University, Lanzhou 730000, China.}
\address[label1]{Institute of Computational Physics and Complex Systems, Lanzhou University, Lanzhou 730000, China} \address[label2]{Key Laboratory for Magnetism and Magnetic Materials of the Ministry of Education, Lanzhou University, Lanzhou 730000, China}
%\fntext[label3]{}

\title{Selfish punishment with avoiding mechanism can alleviate both first-order and second-order social dilemma}

\begin{abstract}
Punishment, especially selfish punishment, has recently been identified as a potent promoter in sustaining or even enhancing the cooperation among unrelated individuals. However, without other key mechanisms, the first-order social dilemma and second-order social dilemma are still two enduring conundrums in biology and the social sciences even with the presence of punishment. In the present study, we investigate a spatial evolutionary four-strategy prisoner's dilemma game model with avoiding mechanism, where the four strategies are cooperation, defection, altruistic and selfish punishment. By introducing the low level of random mutation of strategies, we demonstrate that the presence of selfish punishment with avoiding mechanism can alleviate the two kinds of social dilemmas for various parametrizations. In addition, we propose an extended pair approximation method, whose solutions can essentially estimate the dynamical behaviors and final evolutionary frequencies of the four strategies. At last, considering the analogy between our model and the classical Lotka-Volterra system, we introduce interaction webs based on the spatial replicator dynamics and the transformed payoff matrix to qualitatively characterize the emergent co-exist strategy phases, and its validity are supported by extensive simulations.
\end{abstract}


\end{frontmatter}

% \linenumbers

\section{Introduction}
\label{introduction}
For centuries, the scales of most of human economic communities have expanded dramatically from kin-based work-shops to intensively large-scale cooperative groups in which selfish individuals frequently cooperate with other genetically unrelated ones. Kinds of mechanisms or rules have been developed to explain or support the existence of such cooperative behaviors~(\cite{allfor}). In these range of rules, kin selection is merely applied to small kinship groups (\cite{kin1, kin2}). Direct reciprocity (\cite{direct}) can explain the emergence of cooperation between unrelated individuals or even between members of different species, but it is limited to the repeated encounters between the same two individuals. In the context of indirect reciprocity (\cite{allfor,indirect1, indirect2}), randomly chosen pairwise encounters where two individuals do not have to meet again are admissible. However, it can only promote cooperation on condition with sufficient reputation (\cite{reputation1, reputation2}) that drives this deed. Additionally, network reciprocity (\cite{nowaknet,allfor,cooperation2}) and social diversity (\cite{cooperation3}) are only established in the population that is not well-mixed, i.e., their operations rely heavily on the hierarchical structure of populations. The significance of migration for the emergence and persistence of cooperation has also been highlighted by the previous studies (\cite{migration1,migration2}). Nevertheless, in real life, the cost of migration may be very high, and the information about the destinations may also be insufficient and limited~(\cite{migrationcost1,migrationcost2,migrationcost3,migrationcost4}). For these reasons, punishment turns out to be a key role in sustaining the cooperation as strangers frequently engage in interest transactions in large-scale institutions~(\cite{largep1, largep2}).

Furthermore, altruistic punishment has been used as a paradigm to promote cooperation in large populations consisting of selfish unrelated or faceless individuals~(\cite{Alpunishment, Alpunishment6}). However, it is less likely to become a robust strategy (\cite{nonstable}) owing to the extra expenses for the cost to punish  defectors. Only recently, a few works have suggested another way that altruism may be maintained by the defectors though punishing other defectors, known as selfish punishment~(\cite{selfishpunish,spatial1,selfishpunish2}). The concept of selfish punishment was originally suggested  by an empirical experiment on humans demonstrating that individuals most likely to punish other defectors themselves are most tempted to defect~(\cite{selfishpunish}). This experiment actually implies that a certain part of defectors prefer to punish other defectors for themselves rather the public welfares in some situations. Combing with reality, it is possible that certain groups of individuals, especially tricky cheaters, can take some steps such as lying to punishers to avoid the sanctions in the presence of communication~(\cite{lie2,lie3,lie4}), exemplified by the proverb `a thief crying ``Stop thief"'. These cheaters use punishment as an evasion, which can be regarded as the alternative form of selfish punishment.
However, to our knowledge, this actual important mechanism has received relatively little attention in evolutionary game theory. Further studies are still necessary.

In order to further explore how this selfish punishment works, we design a model involving  cooperators (C), defectors (D), cooperative punishers (CP, i.e., altruistic punishers), and defective defectors (DP, i.e., selfish punishers) with avoiding mechanism. Differing from the previous models~(\cite{cost1,cost2,cost3,spatial1,largep2,Alpunishment,Alpunishment6}) with respect to punishment, our model is performed in the context of prisoner's dilemma game (PDG) along with a low level of random mutation. In detail, the sanctions from punishment are always considered to be costly~(\cite{cost1,cost2,cost3}). %If found and identified, those performing defection will bear a fine while the punishers must pay the cost of punishment.
Similarly, both defective punishers and cooperative punsihers sanction defectors with a punishment fine at a personal punishment cost in our model, without loss of generality and rationality. Unlike previous studies on the evolution of altruism with punishment (\cite{cost1,cost2,cost3,spatial1,largep2,Alpunishment,Alpunishment6}), we add the avoiding behaviors represented by the identifying probability to defective punishers, who punish not only other defectors but also the ones with the same strategy. %Our research shows that the behaviors of the system for different values of avoiding probability depend heavily on both the temptation to defect and the punishment fine and cost.
Moreover, we propose an extended pair approximation for the time evolution of the four strategies, allowing us to track the dynamics features and stationary states of the system. At last, motivated by the works on ecological interaction networks~(\cite{lv}), we introduce interaction webs to qualitatively understand the stable coexistence and extinction of different strategies.  

It is worth noting that there are two kinds of social dilemmas in the model. One is the conventional social dilemma -- PDG, namely the first-order social dilemma in which the free riders such as defectors can earn higher personal profits than cooperators whereas the well-being of the population depend only on the level of cooperation. The other is the second-order social dilemma, where punishers carry out punishment which reduces their fitnesses relative to those second-order free riders (including pure cooperators) who do not punish~(\cite{spatial1,sdilemmas1,sdilemmas2,sdilemmas3}). In this paper, it will be  demonstrated that the presence of selfish punishment with avoiding mechanism can help the individuals out of the two dilemmas.

\section{Model}
\label{model}
We consider a spatially structured population where each player is fastened on one site of a square lattice of size $N=L\times L$ with periodic boundaries. Each player ($i$) adopts strategy $s_{i}\in\{C,D,CP,DP\}$. Initially, the four strategies (C, D, CP, and DP) are distributed randomly and uniformly over the lattice sites.

In each iteration, each player ($i$) firstly plays the PDG with its four nearest neighbors in addition to itself to accumulate its original overall payoff $P^{o}_{s_{i}}$ without punishment. We have found that the situation where cooperators and altruistic punishers in addition to selfish punishers coexisting stably will not be fulfilled for various parametrizations if self-interaction (that the players can play the game with themselves) is excluded. Without self-interaction, the positive role of selfish punishment on the evolution of cooperation is weakened. We thus introduce the self-interaction into the current model. In line with the definition of PDG, each player gets the \emph{reward} $R$ if both choose to cooperate (C, CP) with each other, or the payoff $P$ if both defect (D, DP). A cooperator or cooperative punisher gets the \emph{sucker's} payoff $S$ against a defector or defective punisher, who gets the \emph{temptation} to defect $T$ in such circumstance. We have checked that none of our findings for $T=b (b>1)$, $R=1$, and $P=S=0$ are essentially changed if we instead set $P=\varepsilon$ where $\varepsilon$ is positive but significantly below unity. For the sake of simplicity, we just use the parametrization $T=b~(b>1)$, $R=1$, and $P=S=0$, which is also commonly adopted in many studies~(\cite{norm1, imitate, matrix3, matrix4}).

Secondly, the punishment is executed, i.e., the payoff $P^{o}_{s_{i}}$ may be modified as the remaining payoff $P^{m}_{s_{i}}$ by subtracting punishment costs and/or punishment fines. In reality, the cheaters in the face of punishers have a strong incentive to hide their identities after a defection so as to escape the punishment, causing information asymmetry. Considering this fact, we make an assumption that the states of  non-cooperative (D and DP) individuals are unobservable to other punishers in our model. Consequently, there are three cases as follows: (i) To each punisher ($s_{i}$= CP or DP), it just selects one target $j$ randomly from the non-cooperative neighbors ($s_{j}$= D or DP, $j\in\Gamma_{i}$ where $\Gamma_{i}$ denotes the set of neighbours of player $i$) to identify the target either successfully (for D) or probably (for DP at a probability $\gamma$). Then the punisher $i$ will impose a fine $\beta$ on the exposed target at a personal cost $\alpha$ if its original payoff is sufficient for punishment ($P^{o}_{s_{i}}>\alpha$), or else it will do nothing. It indicates that defective punishers (DP) can still avoid being punished with probability $1-\gamma$ even though they are  selected. (ii) Correspondingly, the selected non-cooperative player $j$ will be either absolutely sanctioned if $s_{j}$= D or successfully punished with a probability $\gamma$ when $s_{j}$= DP, so that its payoff is reduced by $\beta$. Instead, the unselected ones are capable of escaping  the punishment. (iii) $P^{m}_{C}=P^{o}_{C}$ if $s_{i}$= C. 
A run for punishment over the whole lattice is performed in a random fashion in which each punisher has and only has one chance to punish.

Next, each player $i$ chooses one of its four nearest neighbors at random and imitates the strategy of the chosen co-player $j$ with a probability~(\cite{imitate})
\begin{eqnarray}
W_{s_{i}\rightarrow s_{j}}(P^{m}_{s_{j}}-P^{m}_{s_{i}})=1/\{1+\exp[-(P^{m}_{s_{j}}-P^{m}_{s_{i}})/\kappa]\},
 \label{switch}
\end{eqnarray} where the remaining payoff of $j$ ($P^{m}_{s_{j}}$) are also acquired in the same way mentioned above. $\kappa=0.1$ is a noise parameter describing uncertainty of strategy adoption. Over one whole Monte Carlo step (MCS), all the individuals perform an attempt for strategy updating simultaneously.

The previous studies~(\cite{mix0,cost1}) have stressed that the random mutation can create some extent of strategy-mixing, and the introduction of mutation enable us to yield the deterministic replicator dynamics in the limit of frequent sampling in the absence of mutations in a large population~(\cite{mix2}). Moreover, in the presence of the mutation, even though altruistic punishers and selfish punishers evolving from random mutation frequencies cannot stabilise full cooperation~(\cite{mix0}), they could have more chances to touch the first-order free riders (D) and second-order free riders (C)~(\cite{cost1}) to suppress them. The performance of punishment would be thus enhanced by random mutation. Therefore, following the previous studies~(\cite{cost1,mutation}), random mutation is introduced as a separate process in our model. In detail, each player changes its strategy blindly and independently to one of other three strategies with a probability $\nu$ such that all four states are potentially present in the population at a low frequency. %Actually, the above described strategy evolution process is actually performed with probability $1-\nu$.
We set $\nu=10^{-2}$ throughout this paper.

The simulations are performed for systems with $L\geqslant 200$ in our model. The final densities of all four strategies ($\rho_{s}$) are obtained after at least $1.0\times 10^{4}$ Monte Carlo steps (MCS) to guarantee equilibrium existence, and averaged over 20-50 independent realizations to insure a low variability. The simulation results are also complemented and supported by analytical predications from an extended pair approximation method by taking into account the details of the punishment (see Appendix). To get a detailed qualitative portrait of the stable co-exist phase, we put forward interaction webs in analogy to Lotka-Volterra (LV) networks~(\cite{lv}) being detailed in the following section. %This tool is proved to be useful by the simulation results.

\section{Results}
\label{results}
\begin{figure}[ht!]
\hspace*{-.5cm}
\includegraphics[width=\textwidth]{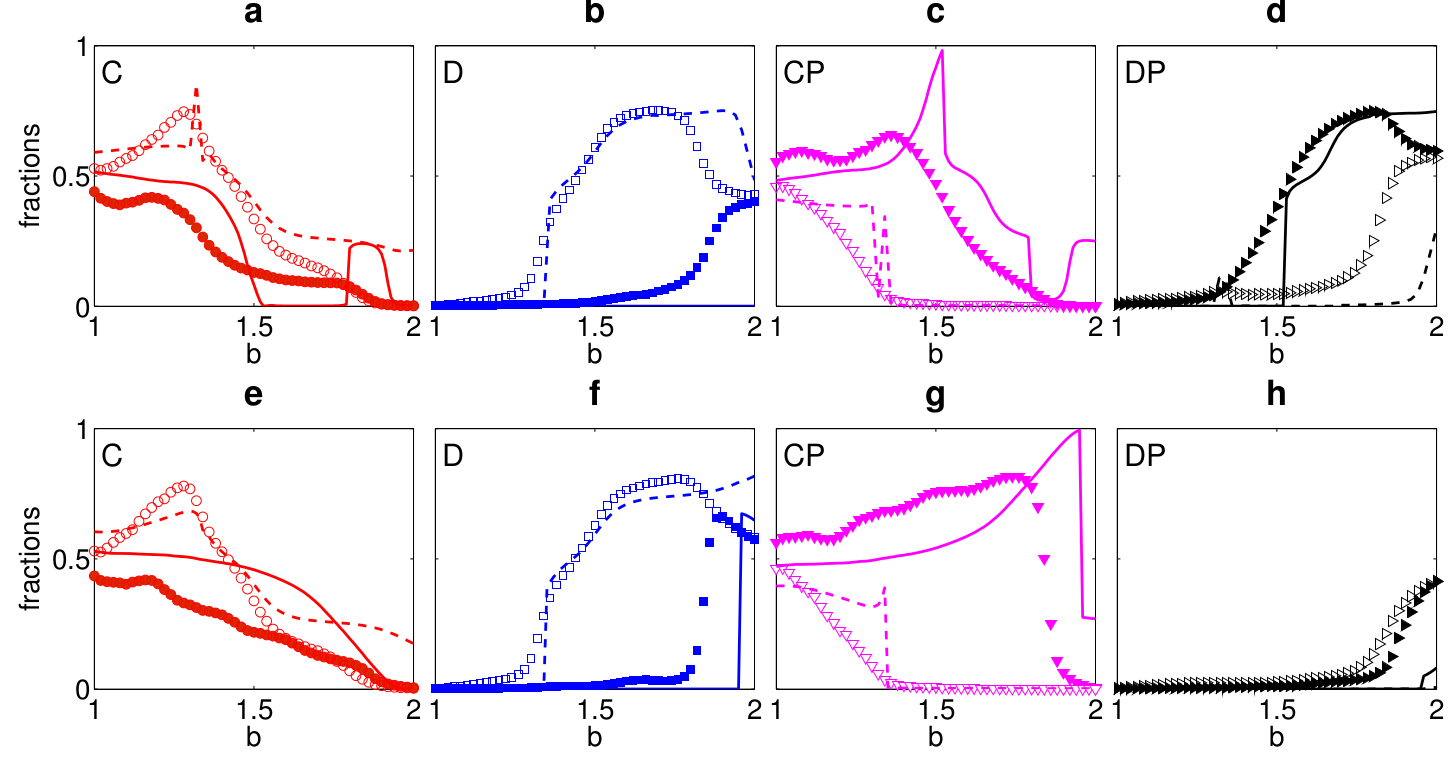}
\caption{The frequencies of four types of strategies as a function of $b$ for $\gamma=0.2$ (top panels (a)-(d)) and $\gamma=0.8$ (bottom panels (e)-(h)), where $\gamma$ denotes the identifying probability that a defective punisher is identified and punished by another punisher. Specifically, solid symbols represent the simulation results for $\alpha=0.3$ and $\beta=0.9$ ($\xi>1$), where the corresponding results based on pair approximation (see Appendix A) are denoted by solid lines. Open symbols represent the simulation results for $\alpha=0.6$ and $\beta=0.4$ ($\xi<1$), where the corresponding results based on the pair approximation are denoted by dashed lines.}
\label{b4}
\end{figure}
Differing from previous studies, selfish punishment with avoiding mechanism diversifies the response of the strategies to punishment cost, punishment fine, and temptation to defect (see Fig.~\ref{b4}). For convenience, we firstly denote punishment coefficient $\xi$ by the ratio of $\beta$ to $\alpha$ i.e., $\xi=\frac{\beta}{\alpha}$. For $\xi<1$ ($\alpha=0.6$, $\beta=0.4$), the frequencies of C (D) does not decrease (increase) monotonically with $b$, independently of the identifying probability $\gamma$. For example, in the case $\gamma=0.2$, when the value of $b$ is just slightly greater than $1$, the population in steady state is almost composed of cooperators and cooperative punishers because non-cooperative individuals even isolated ones can hardly survive in the population. Cooperative punishers have little chance (i.e., don't need to spend external payoffs) to punish defectors. Therefore the random mutation makes that cooperative punishers no great difference to cooperators such that the density of cooperators cannot reach a much high level. When $b$ is approaching $b^{*}=4/3\simeq1.33$, non-cooperative individuals begin to grow and persist in the population, whereas the cost of punishment on non-cooperative individuals is still not so great that cooperative punishers could compete against cooperators. For $b\approx b^{*}$, the payoffs of non-cooperators (D and DP) staying in small clusters less than three (e.g., in forms of lines, corners or isolated islands) can be written as $P_{DP,D}=(k-1)b^{*}\approx 4$, which is comparable to that of clustered cooperators or cooperative punishers in contact with them. In face of second-order free riders, the cooperative punishers owning non-cooperative neighbors (most of them are defectors) are at a disadvantageous position because they have to pay considerable external cost $\alpha$ to punish those nearest defectors in small clusters. Meanwhile defective punishers have to pay much more than the losses of other non-cooperative individuals in spite of that they have a high chance to escape the punishment ($\xi$ is small). In contrast to clusters of D and DP, clusters of C are robust or even aggressive. Consequently, there are almost isolated defectors and clusters of cooperators accompanied by few punishers (CP and DP) in the population in the equilibrium when $b\approx b^{*}$. As such, the density of cooperators has a chance to grow to high level. As $b$ grows approaching to $2$, the population is dominated by defectors and defective punishers, and cooperators and cooperative punishers go extinct. The punishment on defectors is not likely to happen because the payoffs of defective punishers are nearly zero. As a result, there are no differences between D and DP owing to random mutation, they coexist stably together.

For $\xi>1$ ($\alpha=0.3$, $\beta=0.9$), punishers (CP and DP) behave non-monotonically with $b$, relating to $\gamma$. There exists optimal moderate $b$ that could maximize them. Taking the cooperative punishers as an example, in the case $\gamma=0.2$, there are only cooperators and cooperative punishers in the population in the steady state when $b\approx1$. The reason that $\xi>1$ facilitates the survival of punishers (CP and DP) is due to fact that they don't need to spend too much cost to suppress non-cooperative individuals (D and DP). As $b$ is approaching $b=1.5$, cooperators begin to shrink while both cooperative punishers and defective punishers are growing. When $b\approx1.5$, the defective punishers especially the ones in small clusters (existing in form of lines, corners or isolated islands) of the non-cooperative strategies can firstly prevail because they can obtain large benefits from their nearest non-defective neighbors (C and CP) ($P^{o}_{DP}\geq (k-1)*b> 4$ while $P^{o}_{DP}\geq (k-2)*b< 4$). After punishment, they still own enough incomes (because $P^{m}_{DP}=P^{o}_{DP}-\alpha=\geq (k-1)*b-\alpha \approx 4$) to conquer the domains of cooperators and protect cooperative punishers from the excessive invasion of cooperators. Besides, the clusters of CP can also coexist with those of DP on account of the appropriate value of $b$ (the payoffs of the defective punishers in larger clusters $P^{m}_{DP}=P^{o}_{DP}-\alpha=\geq (k-2)*b-\alpha <4$).
Irrespective of small identifying probability, the defective punishers could be still reduced by being probably targeted and punished by the cooperative punishers. When $b\gg1.5$, the population is once again dominated by either defectors or defective punishers based on the arguments in the last paragraph. Consequently, besides fair amount of defective punishers for $b\approx 1.5$, there are greatest number of cooperative punishers in the population in comparison with that for other values of $b$.

The results in Fig.~\ref{b4} show that the evolution of the population is strongly related to the punishment cost and fine, the identifying probability, and the temptation to defect. However, despite of the identifying probability and punishment cost and fine, non-defective individuals (C and CP) finally go extinct as $b$ is approaching to $2.0$, in contrary to the prevalence of defectors and defective punishers. The phenomena observed above are also supported and corroborated by an extended pair approximation method by considering the detailed rules of the punishment (see Appendix A). The method  is sufficient to provide predictions in accordance with the MC simulations.

\begin{figure*}[ht!]
\centering
\includegraphics[width=\textwidth]{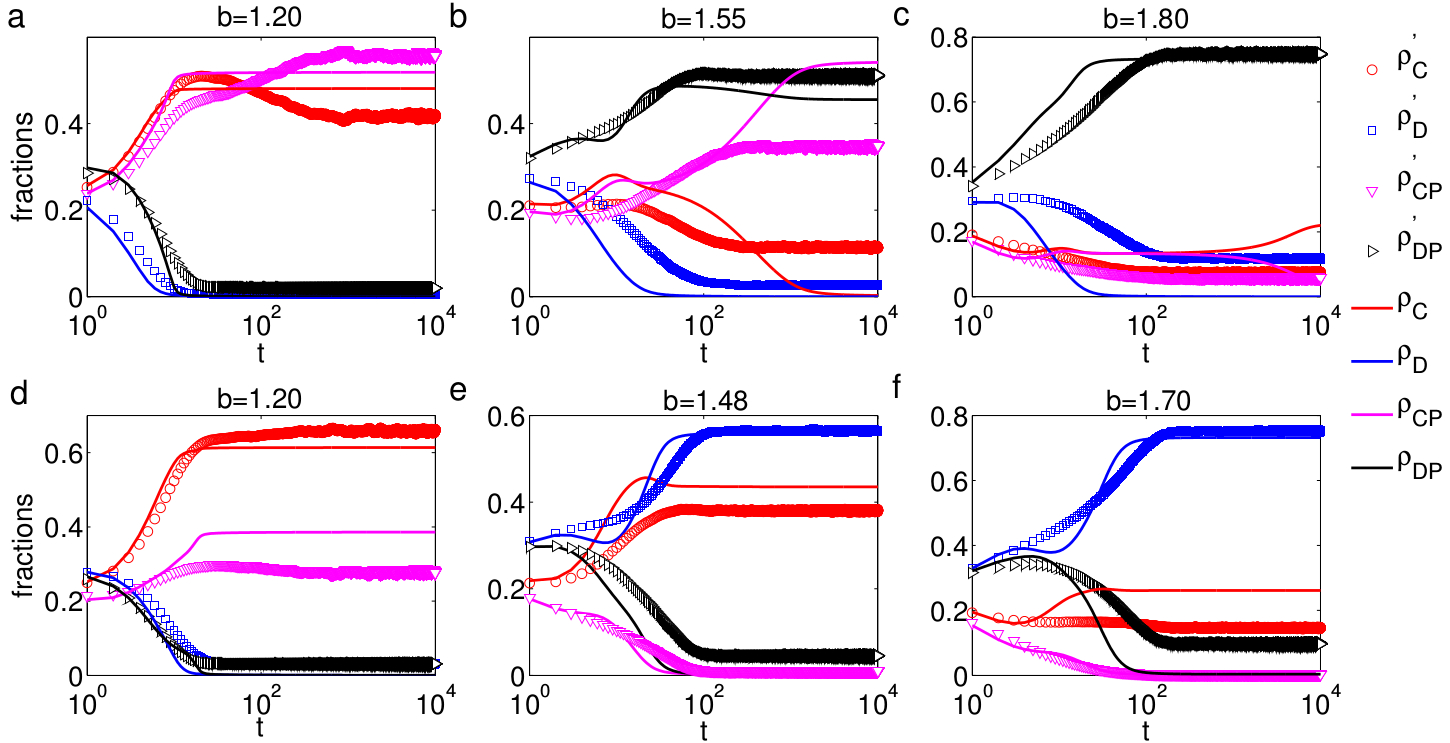}
\caption{Evolution of the four strategies over time for different values of $b$ with $\gamma=0.2$, on condition that $\alpha <\beta$ ($\alpha=0.3$ and $\beta=0.9$ a-c) and $\alpha >\beta$ ($\alpha=0.6$ and $\beta=0.4$ d-f). Open symbols represent simulation results, where solid lines with the same color correspond to solutions of the pair approximation. As displayed, the blue (black) lines are more closed to the corresponding blue (black) symbols than the other lines. (For interpretation of the references to color in this figure caption, the reader is referred to the web version of this article.)}
\label{time2}
\end{figure*}
For a better understanding of the dynamic features of our model, we portray the temporal evolution of the system as appropriate combinations of strategy frequencies in Fig.~\ref{time2} with increasing value of the temptation to defect. It is observed that the mixture of two or more strategies rise simultaneously, while the others fall at the moment. On the one hand, individuals of certain strategies form cooperative islands which are beneficial for them so that they can spread efficiently and nearly conquer the whole population together~(\cite{imitate,cost2,cost3}). For instance, in Fig.~\ref{time2}a and d, cooperators thrive because cooperative punishers can force the non-cooperative ones (D and DP) out on account of small $b$. Meanwhile, cooperative punishers  compensate the cost of punishing non-cooperative individuals by cooperating with other cooperators. As a result, the alliance of both cooperators and cooperative punishers are the final winners. In Fig.~\ref{time2}e, punishers are restricted because they have to pay much more than the losses of non-cooperative individuals (D and DP) to be punished for $\xi>1$. In contrast, both defectors and cooperators can outcompete the punishers because they are not burdened by punishment costs. On the other hand, it also appears that a whole set of combinations of altruism towards members of the strategies performing poorly and spiteful attitudes towards aggressive members can be evolutionarily stable, despite the low fractions of these aligned strategies. For example, in Fig.~\ref{time2}, cooperative punishers can coexist harmoniously (because CP and DP form an anti-coordination game) with defective punishers, i.e. the dominators of the population. Hence cooperators ally with them to avoid excessive exploitations by the dominators, followed by defectors feeding on cooperators. Similarly, C+D+CP, and C+DP in Figs.~\ref{time2}c and f form an alliance to fight against the evasion of the dominant strategies DP and D, respectively. Essentially, inter-group like altruism and invasions of dominant strategies provide a counter-balance for each other. In addition, as soon as the alliances starts to work, the frequencies of strategies change with an almost constant ratio, which have been checked by complementary evaluations.

\begin{figure*}[ht!]
\centering
\includegraphics[width=\textwidth]{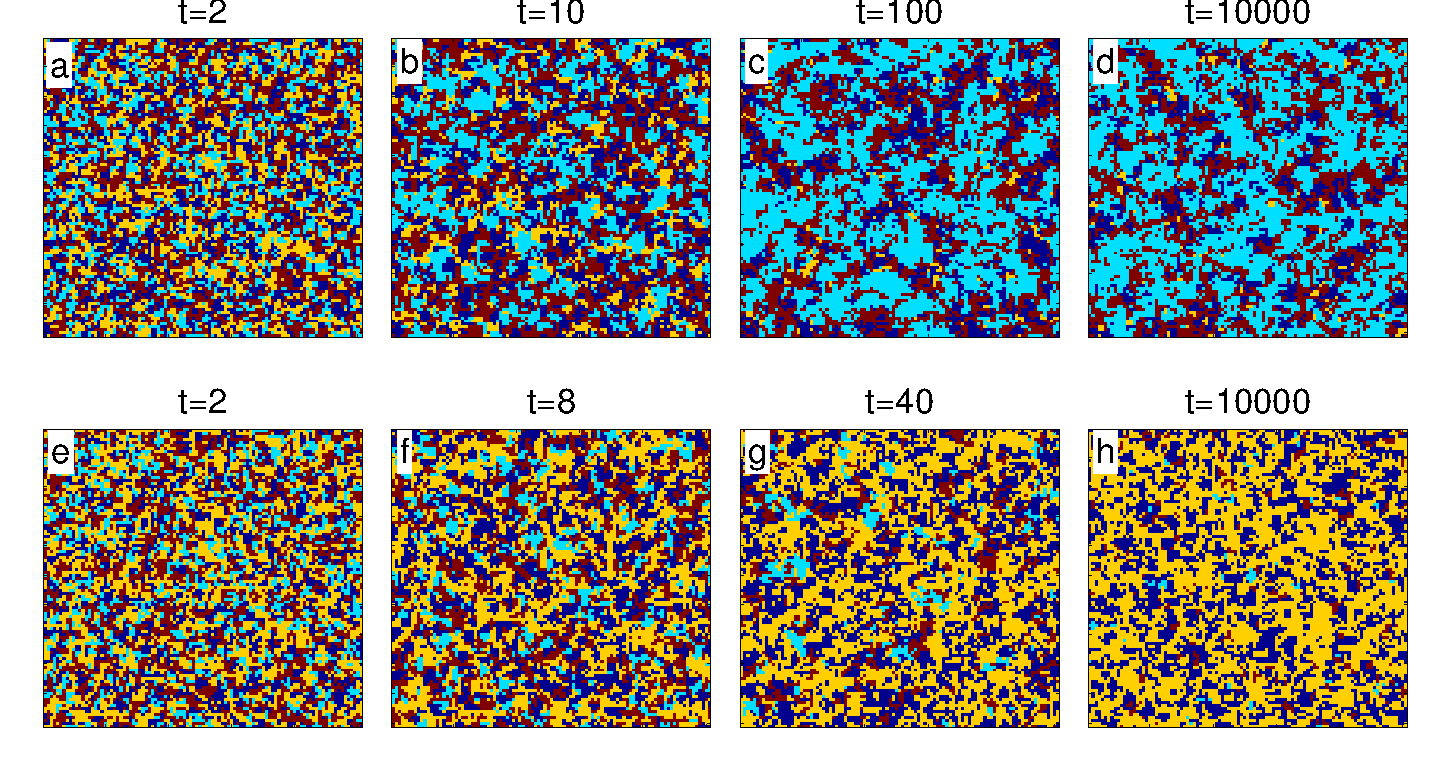}
\caption{Typical snapshots of the simulation grid for $b=1.48$ $\gamma=0.2$ with two groups of punishment cost and punishment fine, $\alpha=0.3$ and $\beta=0.9$ (top panels), $\alpha=0.6$ and $\beta=0.4$ (bottom panels) at different iterations displayed in above figure. Here, $100\times 100$ windows of computer simulations on a $200\times 200$ lattice are shown. Cooperators are represented by dark blue, cooperative punishers by baby blue, defectors by croci and defective punishers by brown.}
\label{spatial}
\end{figure*}
The two types of behaviors observed above can be described as 'alliance phenomena'--two or more strategies form alliances or clusters to either repel the aggressive strategies or win the territorial battles efficiently. It can be intuitively verified in snapshots of the time evolution depicted in Fig.~\ref{spatial}. In the top panels, all clusters of cooperators border that of cooperative punishers in order that they can both avoid exploitations from defective punishers and earn enough payoffs by cooperating with cooperative punishers. In the bottom panels of Fig.~\ref{spatial}, clusters of defective punishers only contact with cooperators besides defective punishers. Depending on sufficient payoffs, they can resist the aggression from defectors to survive.  %get into more resistent clusters of defective punishers (see Fig.~\ref{spatial}a-d) or form mixed islands with the two punishers DP and CP (see Fig.~\ref{spatial}e-h), so as to escape aggression from cooperative punishers or defectors, although sometimes the alliance effect is slight.
The alliance phenomenon is also observed widespread in reality and supported by studies on microbiological social behaviors~(\cite{micro}) and strategic alliance~(\cite{alliance2,alliance1}).

Although there are small deviations, the pair approximation still provides excellent predictions for the final steady frequencies (Fig.~\ref{b4}) and evolutionary processes~(Fig.~\ref{time2}) of the four strategies. Additionally, it can even well reproduce extinction points (Fig.~\ref{b4}). Actually, the deviations between the simulations and the approximations rise from the the fact the individuals of the same or different types gather into clusters together (Fig.~\ref{spatial}), especially cooperators and cooperative punishers (Fig.~\ref{spatial}a-c). It deviates from the assumption of infinite well-mixed populations of pair approximation, which makes the approximation work relatively poor in estimating the dynamic behaviors of cooperators and cooperative punishers.

\begin{figure*}[ht!]
\centering
\includegraphics[width=\textwidth,height=8cm]{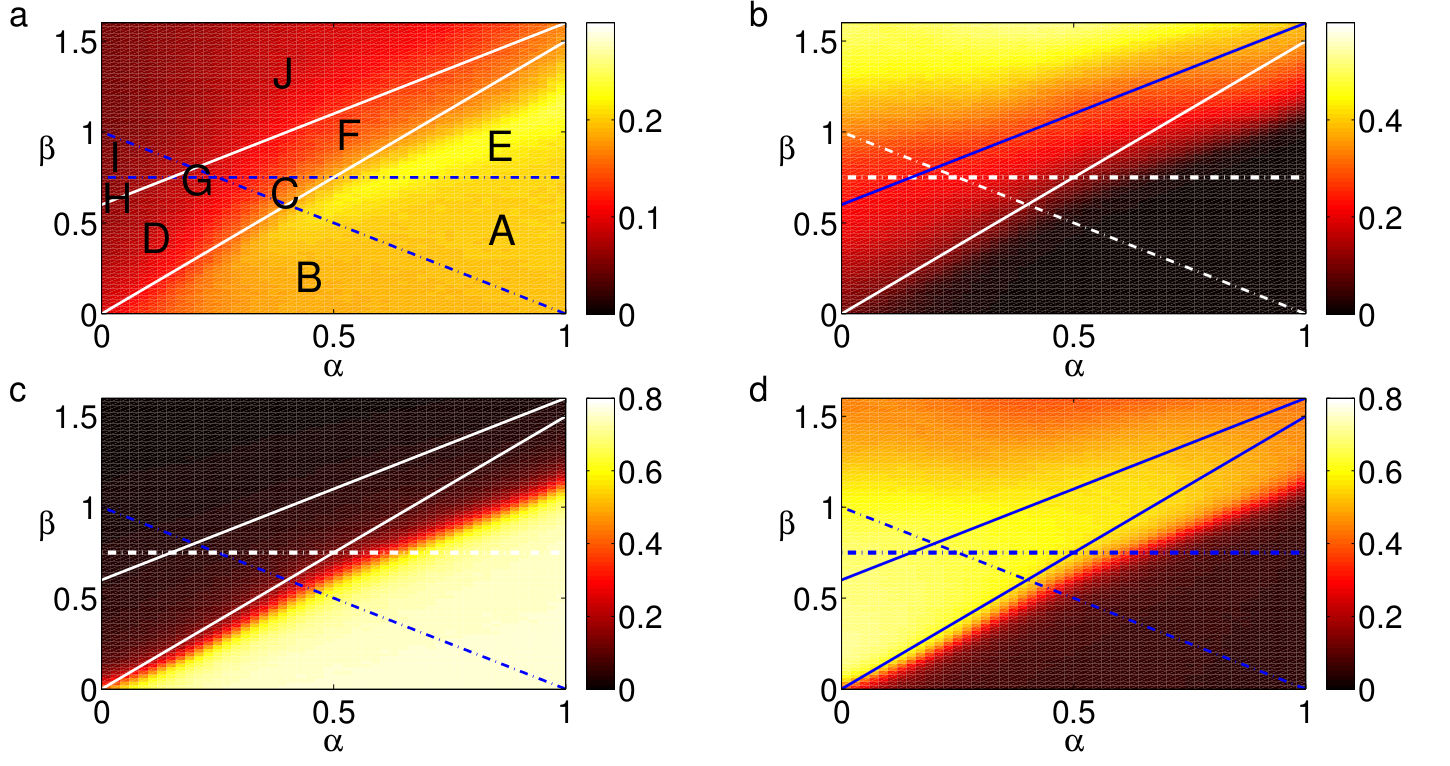}
\caption{Final stationary distributions of the four strategies cooperators (a), cooperative punishers (b), defectors (c) and defective punishers (d) as a function of $\alpha$ and $\beta$. The parameters are taken as $b=1.6$ and $\gamma=0.2$. The final results are obtained by averaging 20 independent runs. Note that the separation between yellow and red reveals that the fraction of strategies change sharply, but does not represent a phase boundary. The four lines are $\beta=-\alpha+\frac{b-1}{3\gamma}$ (dash dotted line), $\beta=\frac{b-1}{4\gamma}$ (dash dotted line), $\beta=\frac{1+\gamma}{1-\gamma}\alpha$ (solid line), and $\beta=b-1+\alpha$ (solid line). The dash dotted lines give the boundaries between $A\longleftrightarrow B$ and $A\longrightarrow B$ (or $B\longrightarrow A$), and the solid lines for $A\longrightarrow B$ and $B\longrightarrow A$. They divide the whole parameter region into ten subregions as marked in the first subgraph (a). As displayed in the figure, there are no four-strategy phases $C+D+CP+DP$ and $C+CP+D$ but obscure three-strategy phase $C+CP+DP$. %(see Appendix B for more details).
The simulations partially agree the predictions from the method based on the spatial predictor equations.}
\label{m162}
\end{figure*}

In the model, non-defective individuals own the `self-sustaining' advantageous ability of being able to benefit from cooperating with themselves. Meanwhile not all non-cooperative individuals selected and targeted have to be punished because of the prerequisite that the payoffs of its neighboring punishers must be sufficient for an execution. To further investigate the interactions and coexistence between different strategies shown in Fig.~\ref{m162}, we make a reasonable assumption that all non-cooperative ones selected and targeted by their neighboring punishers have to be punished, as well as that non-defective individuals cannot earn extra income by playing with themselves. It is equivalent to the case that the two advantages of non-defective individuals and non-cooperative ones cancel out each other. We thus obtain the normalized payoff matrix $\textbf{A}=[a_{ij}]$ with vanishing diagonals
\begin{eqnarray}
%\textbf{A}=
\bordermatrix{%
& C & CP & D & DP \cr
C   & 0 & 0 & 0  &   \gamma(\alpha+\beta)   \cr
CP  & 0 & 0 & -\alpha  & \gamma\beta \cr
D  & b-1 & b-\beta-1  & 0  & \gamma \alpha + (\gamma-1)\beta \cr
DP  & b-1 & b-\gamma \beta-1 & -\alpha & 0\cr
},
\label{paymatrix}
\end{eqnarray}
 without changing the competitive dynamics of the strategies. However, it is known that pair approximation and standard replicator dynamics are not consistent. In the presence of weak selection, \cite{st1,st2} have developed spatial evolutionary dynamics and ESS conditions on regular graphs, showing that evolutionary stability on graphs does not imply evolutionary stability in a well-mixed population. In the limit of the weak selection, the spatial evolutionary dynamics have the form of replicator equations with a transformed payoff matrix $\textbf{B}=[b_{ij}]$ where $b_{ij}=\frac{a_{ii}+a_{ij}-a_{jj}-a_{ji}}{k-2}$ originating from pairwise comparison updating~(\cite{st1}). Therefore, we get the new payoff matrix $\textbf{C}=[c_{ij}]$
\begin{eqnarray}
\hspace*{-2.1cm}
%\textbf{C}=
\bordermatrix{%
& C & CP & D & DP \cr
C   & 0 & 0 & \frac{1}{2}(1-b)  &   \frac{1}{2}(1-b)+\frac{3}{2}(\alpha+\beta)\gamma   \cr
CP  & 0 & 0 & \frac{1}{2}(1-b)+\frac{1}{2}(\beta-3\alpha)   & \frac{1}{2}(1-b)+2\beta \gamma \cr
D  & \frac{3}{2}(b-1) & \frac{3}{2}(b-1)+\frac{1}{2}(\alpha-3\beta)  & 0  & \frac{3}{2}(\alpha+\beta)\gamma +\frac{1}{2}(\alpha-3\beta) \cr
DP  & \frac{3}{2}(b-1)-\frac{1}{2}(\alpha+\beta)\gamma  & \frac{3}{2}(b-1)-2\beta\gamma  & \frac{1}{2}(\beta-3\alpha)-\frac{1}{2}(\alpha+\beta)\gamma & 0\cr
};
\label{paymatrix}
\end{eqnarray}
 which is sum of the original payoff matrix $\textbf{A}$ and the transmitted payoff metrix $\textbf{B}$ (i.e., $c_{ij}=a_{ij}+b_{ij}$). Using matrix $\textbf{C}$ to consider the effects of spatial structure of square lattice, we can determine whether a four-strategy phase (four strategies coexist stably together in the population) or three-strategy phase (three of the four strategies coexist stably together in the population) emerges. Accordingly, we have checked that the four-strategy phase is impossible in the population. There exist three-strategy phases at most. For three-strategy phases, $C+CP+D$ and $C+CP+DP$ are impossible, whereas $C+D+DP$ ($CP+D+DP$) exists conditionally~(for more details see figures form Fig.~\ref{four1} to Fig.~\ref{cpddp} in Appendix. B).

\begin{figure}[ht!]
\centering
\includegraphics[width=\textwidth]{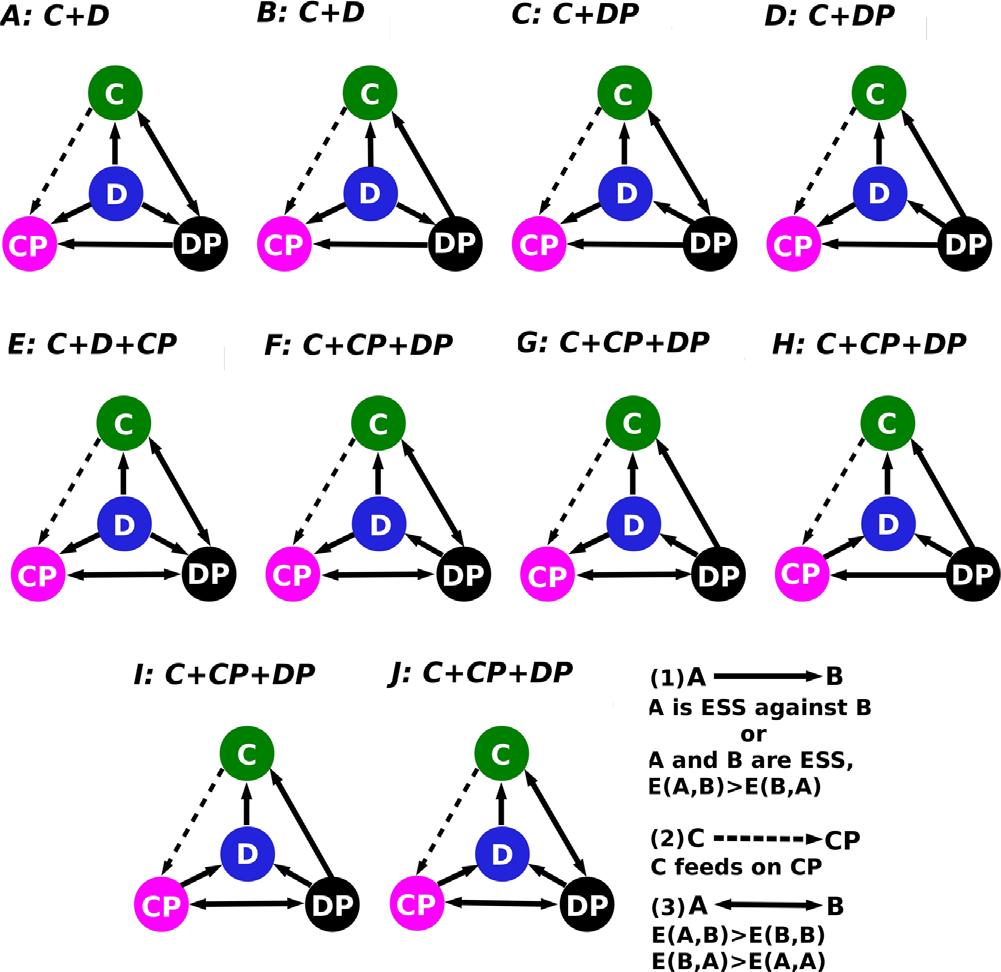}
\caption{The interaction webs for the ten corresponding regions shown in Fig~\ref{m162}, including the co-existing phase induced from the webs. The inducing rules are as follows: (i) On account of `self-sustaining' and 'clustering' advantages of cooperators and cooperative punishers (they can benefit from playing game with themselves and clustering others), they will be "eaten up" only if there are no one-way  arrows pointing other strategies from C or CP and no coexisting edges ($\longleftrightarrow$) connecting them, or else they would still survive in the population if they feed on others or coexist stably with others. (ii) For defectors and defective punishers, they would just exist or thrive if there are no one-way  arrows pointing to them but the one-way arrows from them or the coexisting edges connecting them.} 
\label{web1}
\end{figure}

However, the above complicated quantitative method could not be used to justify the possibilities of two-strategy phases. Through studying the interplay among the four strategies, we notice the similar mechanisms such as competition and predator-prey-like manner~(\cite{prey}) underlying the stationary coexistence of the strategies. The functions of the system is in analogy to LV systems~(\cite{lv}) even though the individuals of different strategies compete indirectly with each other on the basis of their accumulative payoffs rather than the reaction rates. Motivated by these studies, we develop a simple approach to qualitatively understand the stable coexistence and extinction of strategies.
The interactions among the strategies can be well described and further visualized by a fully-connected interaction web in which each vertex (edge/arrow) represents a strategy (the interacting properties of two strategies). Given an example of the case $CP+D$, the form of the new payoff matrix is
$\bordermatrix{%
&   &   \cr
CP  & 0 & \frac{1}{2}(1-b)+\frac{1}{2}(\beta-3\alpha) \cr
D   & \frac{3}{2}(b-1)+\frac{1}{2}(\alpha-3\beta)  & 0 \cr
}$. For $\beta <\frac{\alpha}{3}+(b-1)$, strategy $D$ is absolutely an ESS against $C$, which can be expressed as $D \longrightarrow CP$ in the interaction webs. For $\frac{\alpha}{3}+(b-1)<\beta <3\alpha+(b-1)$, it is a coordination game in which who is the ultimate winner depends on that which strategy is more advantageous in the competition. In more detail, CP is the ultimate strategy when $\beta+(b-1) <\beta<3\alpha+ b-1$ (also expressed as $CP \longrightarrow D$ for simplicity because the relationship between the two strategies in the model is similar to the predator-prey-like manner that one strategy is an ESS against another) because $E(CP,D)>E(D,CP)$, otherwise D is the winner ($D \longrightarrow CP$)~(\cite{original}) when $\frac{\alpha}{3}+(b-1) <\beta<\alpha+(b-1)$. Furthermore, $A\longleftrightarrow B$ means that it is an anti-coordination game giving rise to a stable mixture of strategy A and strategy B. Dashed arrows represent the special predator-prey-like manner between cooperators and cooperative punishers in any situation. With this method, we can give the interaction webs (see Fig.~\ref{web1}) for the results presented in Fig.~\ref{m162}, where every two circles and the arrow connecting them represent one basic functional unit.

It turns out that we can identify all the stable co-exist phase portraits of the dynamics governed by the values of punishment cost, punishment fine, identifying probability and temptation to defect, according to  the manner presented in the caption of Fig.~\ref{web1}. Essentially, we just simplify multi-body problem into two-body problem by investigating the strategy pairs and ignoring errors arising from the multi-strategy interactions.  Fortunately, this approximate treatment is supported by coincidence between the simulations in Fig.~\ref{m162} and the predictions from Fig.~\ref{web1}.

Most importantly, both Fig.~\ref{m162} and Fig.~\ref{web1} reflect that the two undesirable social dilemmas, conventional dilemma and the second-order dilemma change significantly by adding selfish punishment with avoiding mechanism. It is also found that behaviors of the population can be mainly distinguished by  $\beta=\frac{1+\gamma}{1-\gamma}\alpha$ that characterizes the interaction between defective punishers and defectors, though defectors are also in the face of punishment of cooperative punishers. For $\beta<\frac{1+\gamma}{1-\gamma}\alpha$ ($D \longrightarrow DP$) (regions A, B, and E in Fig.~\ref{m162}), the selfish and altruistic punishers indeed promote the level of cooperation within a polymorphic equilibrium, but sacrifice themselves (and D is absolutely an ESS against DP for $\beta<\frac{1+3\gamma}{3(1-\gamma)}\alpha$). Because the punishers, especially those defective ones, have to bear larger additional punishment costs relative to what they squeeze from cooperators. Their competitiveness are reduced. By contrast, defectors and cooperators avoid extra costs by punishment efforts. Moreover, cooperators can exploit the defection-suppressing benefits created by the punishers. Consequently, they can survive and prevail, further leading to the growth of defectors, which is also confirmed by the predictions based on interaction webs (Fig.~\ref{web1}). This situation is always referred to as the second-order soial dilemma. %Consequently, it can be observed the stable coexistence of free riders and second-order free riders.
Nonetheless, this is changing with $\beta$ getting larger than $\frac{1+\gamma}{1-\gamma}\alpha$ (regions C, D, F, G, H, I, and J in Fig.~\ref{m162}). Defective punishers do not need to entail much to crowd the defectors out. They can both take advantage of cooperators to exploit enough payoffs and efficiently wriggle out of the punishment owing to small $\gamma$. That can also protect the cooperative punishers from excessive invasion of the second-order free riders such as cooperators. As a result, stable coexistences of cooperative punishers and defective punishers, i.e. altruistic punishment and selfish punishment arise and persist. Meanwhile, conventional (first-order) free riders are nearly eliminated, whereas the cooperators can ally and cooperate with altruistic punishers to survive (also see the arguments for Fig.~\ref{time2} and Fig.~\ref{spatial}). It implies that the two social dilemmas are factually alleviated. Departing from previous studies~(\cite{spatial1,spatial2}), the accurate predictions based on interaction webs (Fig.~\ref{web1}) also suggest that the stable coexistences of altruistic punishers, selfish punishers, and cooperators in our model are independent of the spatial neighborhood of the population.

\begin{figure*}[ht!]
\centering
\includegraphics[width=\textwidth]{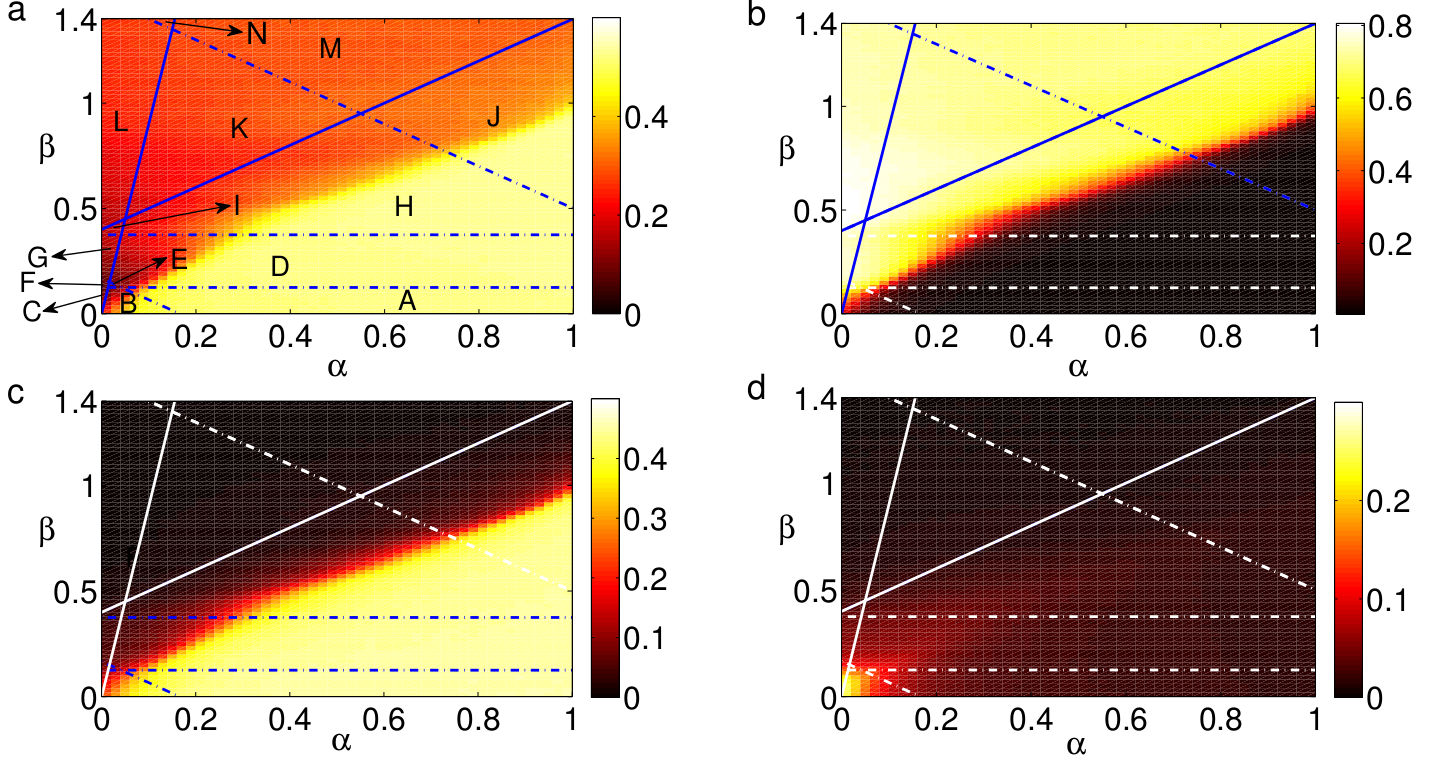}
\caption{Final stationary distributions of the four strategies cooperators (a), cooperative punishers (b), defectors (c) and defective punishers (d) as a function of $\alpha$ and $\beta$. The parameters are taken as $b=1.4$ and $\gamma=0.8$. The final results are obtained by averaging 20 independent runs. Note that the separation between yellow and red indicates that the fractions of strategies change sharply, but does not represent a phase boundary. The dash dotted lines form bottom to top are  $\beta=-\alpha+\frac{b-1}{3\gamma}$, $\beta=\frac{b-1}{4\gamma}$, $\beta=\frac{3(b-1)}{4\gamma}$, and $\beta=-\alpha+\frac{3(b-1)}{\gamma}$. The solid lines from bottom to top are $\beta=\frac{1+\gamma}{1-\gamma}\alpha$ and $\beta=b-1+\alpha$. 
They divide the whole parameter region into fourteen ones as marked in the first subgraph (a). As displayed in the figure, four-strategy phase $C+D+CP+DP$ and three-strategy phase $C+CP+D$ are impossible, but obscure three-strategy phase $C+CP+DP$ can appear.} 
\label{148}
\end{figure*}
As $\gamma$ increases to $0.8$, the interaction between cooperative punishers and defectors (identified by  $\beta=b-1+\alpha$ plotted as dotted lines in Fig.~\ref{148}) plays a decisive role (see Fig~\ref{148}). High identifying probability makes defective punishers no great difference to defectors, resulting in that more selfish punishers will be punished especially for $\beta<b-1+\alpha$. The punishment targeting free riders and the exploitations of cooperators from them are also naturally weakened. As a result, the second social dilemma is still remained (Fig~\ref{148}) for $\beta<b-1+\alpha$ (regions A-J in Fig.~\ref{148}). In the case where $\beta>b-1+\alpha$ (regions K-N in Fig.~\ref{148}), cooperative punishers become superior to defectors ($CP \longrightarrow D$), while cooperators can outcompete cooperative punishers. As a consequence, the mixture of altruistic punishers and cooperators is maintained, suggesting that the second-order dilemma can be resolved. Similarly, we apply the interaction webs to the case where $\gamma=0.8$ to validate the robustness of this method (see Fig~\ref{web2} in Appendix C). It is still valid in most subregions except several ones near the junction of the line $\beta=\frac{1+\gamma}{1-\gamma}\alpha$ estimating the interaction properties between D and DP, the line $\beta=-\alpha+\frac{b-1}{3\gamma}$ for C and DP, and the line $\beta=\frac{3(b-1)}{4\gamma}$ for CP and DP. The reason for the failure of the interaction webs is that the alliance phenomenon of the strategies predicted to be extinct (CP and DP) takes effect. The individuals of the weak strategies (CP and DP) incline to cluster themselves~(\cite{cost1,cost2}) or contact with the more resistant co-exist partners so as to assert against invasion from the aggressive ones (C or D). Nevertheless, the stabilizations of cooperation is established despite the changes of $\alpha$ and $\beta$ (see Fig.~\ref{148}a).

Like what obtained by~\cite{prsb}, both cooperation and altruism can be guaranteed simultaneously in a crowd of faceless individuals for a wide parameter range (just $\beta\gtrsim \alpha$) in our model, by just introducing selfish punishment with avoiding mechanism and without the need for other mechanisms such as reputation. Instead of the results in ~\cite{prsb}, the punishment with avoiding mechanism allows the CP strategy to capture more individuals than that of the second-order free riders (C) for $\beta\gtrsim \alpha$,  where defective punishers thrive as well for small $\gamma$.  Whereas the coexistence of cooperators and defectors is still possible for $\beta\lesssim \alpha$, which is similar to the results  of~(\cite{cost1,cost2}) and contrast to that of~(\cite{prsb}). Differing from the related works~(\cite{prsb,cost1,cost2}), the system converges to a bistable or even multistable state except a homogeneous phase through out the whole parameter ranges in our model. However, our study share with the related researches~(\cite{cost1,cost2,cost3,prsb}) the common feature that increasing punishment fine over punishment cost can facilitate the abundance of punishers especially altruistic ones.

\section{Discussion}
\label{discussion}
In order to deeply explore the impact of the selfish punishment, we have proposed a spatial evolutionary four-strategy prisoner's dilemma game model involving cooperators, defectors, altruistic punishment, and selfish punishment accounting for avoiding behaviors. Meanwhile, we introduced a low level of random strategy mutation into the evolutionary process. Unlike the monotonic changes of strategies revealed by previous studies~(\cite{monotonic1,monotonic2}), we observed the diverse responses of the strategies to changes of punishment cost and fine, identifying probability, and the temptation to defect. It can be found optimal levels of various types of individuals favoured by the different mediate values of $b$. Afterwards, we employ an extended pair approximation method involving four strategies and considering the details of the punishment to estimate the dynamical behaviors and final evolutionary frequencies of the strategies.  Surprisingly, the predications of the method are in well agreement with the simulations. At he same time, in our model, we observed the `alliance phenomena' -- two or more kinds of individuals form alliances or clusters being beneficial for them, so as to efficiently either repel the dominant strategies or even win the territorial battles. At last, motivated by the works on ecological interaction networks, we introduced the interaction webs in analogy to LV networks~(\cite{lv}) to qualitatively address the stable coexistences and extinctions of strategies in terms of new payoff matrix considering the effects of spatial structure of square lattice. The good performance of the interaction webs revealed that the effect of the selfish punishment on the evolution of altruism depends on the punishment cost and fine, the identifying probability, instead of the temptation to defect (~\cite{nowaknet,net3,net2,spatial2}).

The vital function of selfish punishment on the evolution of altruism has been noticed and primarily investigated in \cite{spatial1, selfishpunish2}. In these previous researches it is difficult to sustain or even promote cooperation and altruism simultaneously, particularly second-order altruism (i.e CP) and the selfish punishment itself. Interestingly, in our results, addition of the selfish punishers not only trend to encourage the prevalence of the both punishers (CP and DP) but also ensure the survival of cooperators (C) as long as $\beta>\frac{1+\gamma}{1-\gamma}\alpha$ for small identifying probability. Alternatively, for $\beta>b-1+\alpha$, although the selfish punishers with weak avoiding ability (large $\gamma$) are nearly eliminated by altruistic punishers, the system still arrives at the absorbing C+CP state i.e., cooperators and altruistic punishers coexist in harmony in the population. Nevertheless, in either case, the cooperation in the population is maintained. We thus conclude from the research results in this paper that the selfish punishment with avoiding mechanism can alleviate the two types of social dilemmas at the same time under certain parameter conditions. Moreover, this reveals that defectors are more likely to be punished by other defectors, which can be taken as a complementarity to what stressed by \cite{nowak},

As we known, humans have maintained cooperation in scales ranging from kin-based hunter-gather bands to large modern states over the last 50000 years, accompanied by cheating and altruistic punishment in the form of regulations or codes~(\cite{history}). Theoretically, altruistic punishers suffer a considerable cost from pure cooperation, so they tend to go extinct. Nevertheless, our findings confirm that the selfish punishment with avoiding mechanism can exempted the second-order altruists from this dilemma. In the perspective of evolutionary biology theory, it indicates that maybe the selfish punishment and altruistic punishment evolved at approximately the same time in human history. The present study thus provide a new insight into understanding the origin and persistence of altruism and selfish punishment in social life.

\section*{Acknowledgments}
This work was supported by the National Natural Science Foundation of China (Grant No. 11135001), and by the Fundamental Research Funds for the Central Universities (Grants No. lzujbky-2014-28).

\section*{Appendix A: Extended pair approximation}
\label{apa}
Let $\rho_{s_{x}s_{y}}$ ($s_{x}, s_{y}\in\{C, D, CP, DP\}$) denote the frequency of strategy pairs e.g. $C-C$, $C-D$, $D-DP$ and so on. Each individual owns $k=4$ nearest neighbors, thus there are $12$ types of strategy pairs. $q_{s_{x}|s_{y}}= \rho_{s_{x}s_{y}}/\rho_{s_{y}}$ specifies the conditional probability to find an imitator owning strategy $s_{x}$ given a target neighbor adopting strategy $s_{y}$. The main idea of pair approximation is to describe the dynamics of strategy pairs, meaning that everything has to be expressed in terms of configurations no more complex than pairs. Therefore, on the basis of the compatibility condition ($\rho_{s_{x}}=\sum_{s_{y}}\rho_{s_{x}s_{y}}$), the symmetry condition ($\rho_{s_{x}s_{y}}=\rho_{s_{y}s_{x}}$), and closure conditions ($\sum\limits_{s_{x},s_{y}} \rho_{s_{x}s_{y}}=1$), we choose to pay our attention on nine strategy pairs $C-C$, $C-D$, $C-DP$, $C-CP$, $D-D$, $D-CP$, $D-DP$, $CP-CP$, $CP-DP$. We treat them as the variables for tracing the behaviors of the population.

During the course of the PDG game, changes of strategy frequencies take place only when a target player $x$ switch its strategy to the different strategy of a referencing neighbor $y$. We consider a configuration where an imitator playing strategy $s_{x}$ against the neighboring player with strategy $s_{y}$. Let $k_{C}$, $k_{D}$, $k_{CP}$, and $k_{DP}$ denote the number of cooperators, defectors, cooperative, and defective punishers in imitator's neighborhood on a square lattice, where $k=k_{C}+k_{D}+k_{CP}+k_{DP}$. The frequency of such configuration is
\begin{eqnarray}
T^{k}_{k_{C},k_{D},k_{CP}}q^{k_{C}}_{C|s_{x}}q^{k_{D}}_{D|s_{x}}q^{k_{CP}}_{CP|s_{x}}q^{k-k_{C}-k_{D}-k_{CP}}_{DP|s_{x}}.
\label{config1}
\end{eqnarray} Here, $T^{k}_{k_{C},k_{D},k_{CP}}$ represents the coefficient in the four type expansion, i.e., $T^{k}_{k_{C},k_{D},k_{CP}}=\frac{k!}{k_{C}!k_{D}!k_{CP}!(k-k_{C}-k_{D}-k_{CP})!}$.

The probability of the configuration that the target neighbor $y$ has $k^{'}_{C}$ cooperators, $k^{'}_{D}$ defectors, $k^{'}_{CP}$ cooperative punishers, and $k^{'}_{DP}$ defective punishers among the $k-1$ remaining neighbors including the imitator $x$ is
\begin{eqnarray}
T^{k-1}_{k^{'}_{C},k^{'}_{D},k^{'}_{CP}}q^{k^{'}_{C}}_{C|s_{y}s_{x}}q^{k^{'}_{D}}_{D|s_{y}s_{x}}q^{k^{'}_{CP}}_{CP|s_{y}s_{x}}q^{k-k^{'}_{C}-k^{'}_{D}-k^{'}_{CP}-1}_{DP|s_{y}s_{x}}.
\label{config2}
\end{eqnarray} $q_{s|s_{y}s_{x}}$ ($s\in\{C, D, CP, DP\}$) gives the conditional probability that one player next to the strategy pair $s_{y}-s_{x}$ owns strategy $s$. Furthermore, the triplet configuration is approximated by the assembles of the configurations that are not more complex than pairs ($q_{s|s_{y}s_{x}}\approx q_{s|s_{y}}$) to insure the a moment closure of Eqs.~(\ref{config1}) and (\ref{config2}). Combining with the expression of conditional probability, Eqs~(\ref{config1}) and (\ref{config2}) can be reduced to
\begin{equation}
T^{k}_{k_{C},k_{D},k_{CP}}\Omega_{s_{x}}(k,k_{C},k_{D},k_{CP})\quad  \text{and} \quad
T^{k-1}_{k^{'}_{C},k^{'}_{D},k^{'}_{CP}}\Omega_{s_{y}}(k-1,k^{'}_{C},k^{'}_{D},k^{'}_{CP}),
\label{configl}
\end{equation} where $\Omega_{s_{x}}(k,k_{C},k_{D},k_{CP})=
\frac{\rho^{k_{C}}_{Cs_{x}}\rho^{k_{D}}_{Ds_{x}}\rho^{k_{CP}}_{CPs_{x}}\rho^{k-k_{C}-k_{D}-k_{CP}}_{DPs_{x}}}{\rho^{k}_{s_{x}}}$ and $\Omega_{s_{y}}(k-1,k_{C},k_{D},k_{CP})=
\frac{\rho^{k^{'}_{C}}_{Cs_{x}}\rho^{k^{'}_{D}}_{Ds_{x}}\rho^{k^{'}_{CP}}_{CPs_{x}}\rho^{k-k^{'}_{C}-k^{'}_{D}-k^{'}_{CP}-1}_{DPs_{x}}}{\rho^{k-1}_{s_{y}}}$. Eq.~(\ref{configl}) gives the probability of existence of a configuration surrounding imitator $x$ and target player $y$ in terms of the link densities and strategy frequencies.

By neglecting the neighborhood of (values of $k_{C}$, $k_{D}$ $k_{CP}$, and $k_{DP}$) the imitator, the probability that state of player $x$ switches from $s_{x}$ to $s_{y}$ can be written as
\begin{eqnarray}
Tr^{s_{x}\rightarrow s_{y}}_{s^{'}_{x}s^{'}_{y}}=\frac{k_{s_{y}}}{k}\sum\limits^{k-1}\limits_{k^{'}_{C},k^{'}_{D},k^{'}_{CP}}T^{k-1}_{k^{'}_{C},k^{'}_{D},k^{'}_{CP}}\Omega_{s_{x}}(k-1,k^{'}_{C},k^{'}_{D},k^{'}_{CP})W_{s_{x}\rightarrow s_{y}}(P^{m}_{s_{y}}-P^{m}_{s_{x}}).
\label{change1}
\end{eqnarray}
That leads to a corresponding changes in the number of the strategy pairs (such as pair $s^{'}_{x}-s^{'}_{y}$, where $s^{'}_{x},s^{'}_{y}\in\{C, D, CP DP\}$). We let $\Delta n^{s_{x}\rightarrow s_{y}}_{s^{'}_{x},s^{'}_{y}}(k_{C}, k_{D}, k_{CP})$ denote this change in the number of $s^{'}_{x}-s^{'}_{y}$ pair due to the switching event from $s_{x}$ to $s_{y}$ for given by $k_{C}$, $k_{D}$, and $k_{CP}$. Subsequently taking into account all the possible neighborhood of the imitator $x$, we obtain the final expressions of the overall changes in frequency of strategy pair $s^{'}_{x}-s^{'}_{y}$:
\begin{eqnarray}
\dot{\rho}_{s^{'}_{x}s^{'}_{y}}=\frac{1}{k}\sum\limits_{\substack{s_{x},s_{y}\\k_{C},k_{D},k_{CP}}}T^{k}_{k_{C},k_{D},k_{CP}}\Omega_{s_{x}}(k,k_{C},k_{D},k_{CP})\Delta n^{s_{x}\rightarrow s_{y}}_{s^{'}_{x},s^{'}_{y}}(k_{C}, k_{D}, k_{CP})Tr^{s_{x}\rightarrow s_{y}}_{s^{'}_{x}s^{'}_{y}}
\label{change2}
\end{eqnarray} in the limit of large population sizes $N\rightarrow \infty$~(\cite{pair1}).

Herein $P^{m}_{x}(k_{C},k_{D},k_{CP},k_{DP})$ are accumulated according to the manner described in Sec.~\ref{model}.
For cooperators,
\begin{eqnarray}
P^{m}_{C}(k_{C},k_{D},k_{CP},k_{DP}) & = & k_{C}+k_{CP}+1.0.  \\ \notag
\label{payoff1}
\end{eqnarray}
However, for punishers and defectors, we must consider the situation whether the player and its punishing neighbors are capable of punishing its non-cooperative individuals. Correspondingly, there are three cases. (i) For cooperative punishers,
\begin{eqnarray}
P^{m}_{CP}(k_{C},k_{D},k_{CP},k_{DP}) & = &  k_{C}+k_{CP}+1.0, \quad \text{if} \quad k_{C}+k_{CP}+1.0<\alpha \\ \notag
P^{m}_{CP}(k_{C},k_{D},k_{CP},k_{DP}) & = & k_{C}+k_{CP}+1.0-\frac{k_{D}+\gamma k_{DP}}{k_{D}+k_{DP}}\alpha \quad \text{if} \quad k_{C}+k_{CP}+1.0\geqslant \alpha. \notag
\label{paycp}
\end{eqnarray}
(ii)For defectors,
\begin{eqnarray}
P^{m}_{D}(k_{C},k_{D},k_{CP},k_{DP}) & = & (k_{C}+k_{CP})b \\ \notag & &  \text{if} \left\{
\begin{array}{cccc}
k(q_{C|CP}+q_{CP|CP})+1.0<\alpha \\
k(q_{C|DP}+q_{CP|DP})b<\alpha
\end{array}
\right. \\ \notag
P^{m}_{D}(k_{C},k_{D},k_{CP},k_{DP})  & = & (k_{C}+k_{CP})b-k_{CP}S_{D,CP} \\ \notag & &  \text{if} \left\{
\begin{array}{cccc}
k(q_{C|CP}+q_{CP|CP})+1.0\geqslant \alpha \\
k(q_{C|DP}+q_{CP|DP})b<\alpha
\end{array}
\right. \\ \notag
P^{m}_{D}(k_{C},k_{D},k_{CP},k_{DP})  & = & (k_{C}+k_{CP})b-k_{DP}S_{D,DP} \\ \notag & & \text{if} \left\{
\begin{array}{cccc}
k(q_{C|CP}+q_{CP|CP})+1.0<\alpha\\
k(q_{C|DP}+q_{CP|DP})b\geqslant \alpha
\end{array}
\right.\\ \notag
P^{m}_{D}(k_{C},k_{D},k_{CP},k_{DP})  & = & (k_{C}+k_{CP})b-k_{CP}S_{D,CP}-k_{DP}S_{D,DP} \\ \notag & &  \text{if} \left\{
\begin{array}{cccc}
k(q_{C|CP}+q_{CP|CP})+1.0\geqslant \alpha\\
k(q_{C|DP}+q_{CP|DP})b\geqslant \alpha
\end{array}
\right.
\label{payd}
\end{eqnarray}
(iii) For defective punishers, at first, the original payoff can be obtained according to the following equations:\\
\begin{eqnarray}
P^{o}_{DP}(k_{C},k_{D},k_{CP},k_{DP}) & = & (k_{C}+k_{CP})b \quad \text{if} \quad (k_{C}+k_{CP})b< \alpha ,\\ \notag
P^{o}_{DP}(k_{C},k_{D},k_{CP},k_{DP}) & = &  (k_{C}+k_{CP})b-\frac{k_{D}+\gamma k_{DP}}{k_{D}+k_{DP}}\alpha \quad  \text{if}\quad (k_{C}+k_{CP})b\geqslant \alpha .
\end{eqnarray}
Then the remaining payoffs $P^{m}_{DP}(k_{C},k_{D},k_{CP},k_{DP})$ are obtained following the similar manner as in Eq.~(\ref{payd}). $S_{D,CP}$ and $S_{D,DP}$ ($S_{DP,CP}$ and $S_{DP,DP}$) represents the punishment from one of defector's (defective punisher's) neighbors in state of cooperative punishment and defective punishment, respectively. The probability that a cooperative (defective) punisher selects one defector randomly from its non-cooperative neighbors (D, DP) is $\frac{1}{kq_{D|CP}+kq_{DP|CP}}$ ($\frac{1}{kq_{D|DP}+kq_{DP|DP}}$) through ignoring the impact of loops and the configurations that are more complex than pairs. Consequently, we have $S_{D,CP}=\frac{\beta}{k(q_{D|CP}+q_{DP|CP})}=\frac{\beta}{k(\frac{\rho_{DCP}}{\rho_{CP}}+\frac{\rho_{DPCP}}{\rho_{CP}})}=\frac{\rho_{CP} \beta}{k(\rho_{DCP}+\rho_{DPCP})}$ and $S_{D,DP}=\frac{\beta}{k(q_{D|DP}+q_{DP|DP})}=\frac{\beta}{k(\frac{\rho_{DDP}}{\rho_{DP}}+\frac{\rho_{DPDP}}{\rho_{DP}})}=\frac{\rho_{DP} \beta}{k(\rho_{DDP}+\rho_{DPDP})}$. $S_{DP,CP}=\gamma S_{D,CP}$ and $S_{DP,DP}=\gamma S_{D,DP}$. $P^{m}_{y}(k^{'}_{C},k^{'}_{D},k^{'}_{CP},k^{'}_{DP})$ can be accumulated in the same way.

Given an initial well-mixed condition that $\rho_{s_{x}s_{y}}(0)=\frac{1}{4^{2}}$ and $\rho_{s}(0)=\frac{1}{4}$ ($s\in\{C, D, CP, DP\}$) satisfying the closure condition, we can trace the dynamic changes of the frequencies of strategy pairs according to Eq.~(\ref{change2}). The steady state can be solved by setting $\dot{\rho}_{s^{'}_{x}s^{'}_{y}}=0$ or by numerically iterating these equations to a steady state for a long time. Additionally, combining with the random mutation, the evolution of frequencies of the four strategies can be tracked in the following manner:
\begin{eqnarray}
\rho_{s}(t)=\sum\limits_{\substack{s_{x},s_{y};\\s_{x}\text{or}s_{y}=s; s_{x}=s_{y}=s}}\rho_{s_{x}s_{y}}(t)(1-\nu)+\frac{\nu}{3.0}\sum\limits_{s^{'}\neq s_{x}}\rho_{s^{'}}(t).
\label{frequency}
\end{eqnarray}

\section*{Appendix B: Analysis for strategy phases based on spatial replicator equation}
\label{apb}
\begin{figure}[htbp]
\centering
\includegraphics[width=\textwidth,height=14cm]{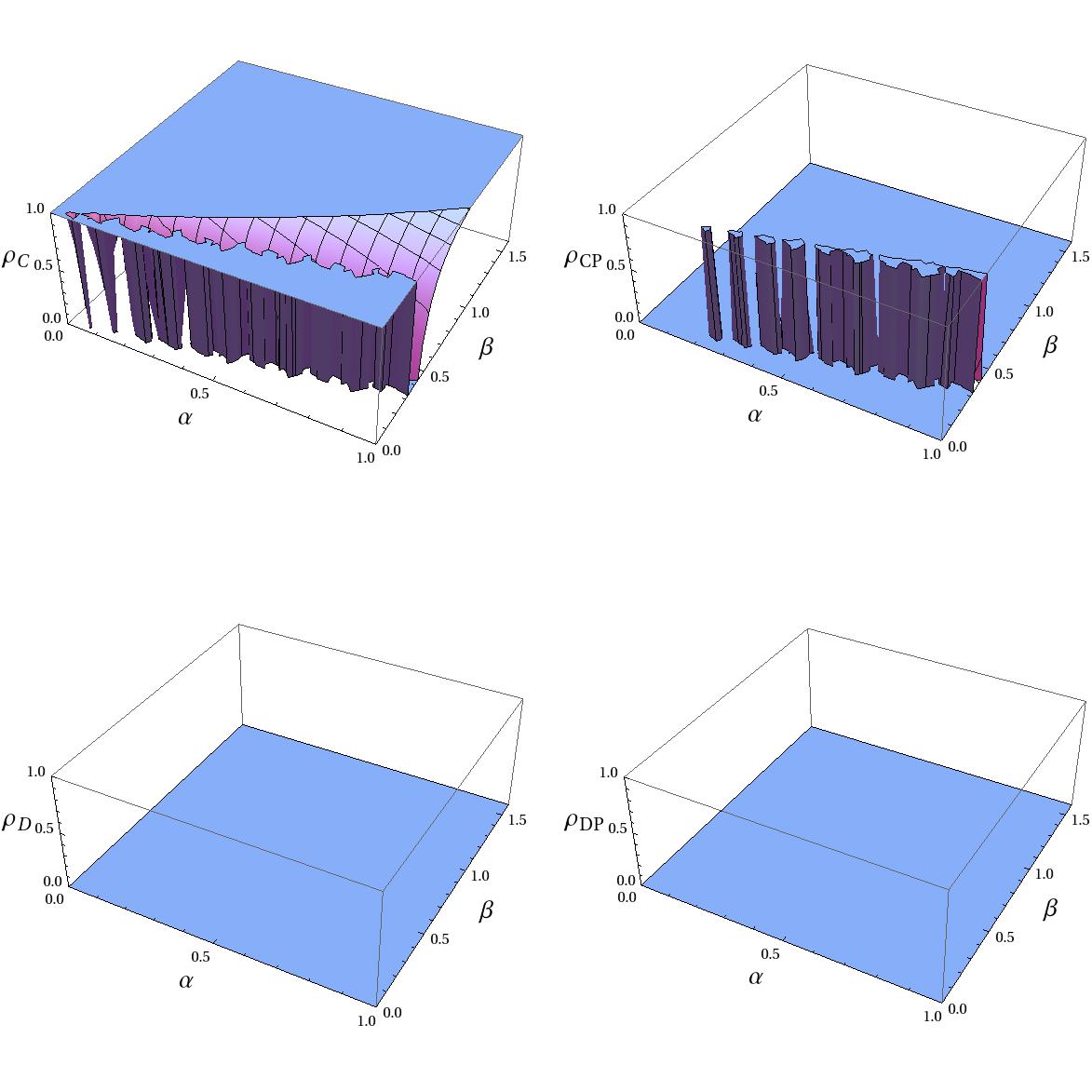}
\caption{The frequencies of four strategies based on spatial replicator equation, as function of $\alpha$ and $\beta$ for $b=1.6$ and $\gamma=0.2$. It can be observed that there are not parameter regions where the relationships $0<\rho_{C}<1$, $0<\rho_{D}<1$, $0<\rho_{CP}< 0$, and $0<\rho_{DP}< 0$ are satisfied at the same time. It indicates that the coexistence of the four strategies is impossible.}
\label{four1}
\end{figure}
\begin{figure}[htbp]
\centering
\includegraphics[width=\textwidth,height=14cm]{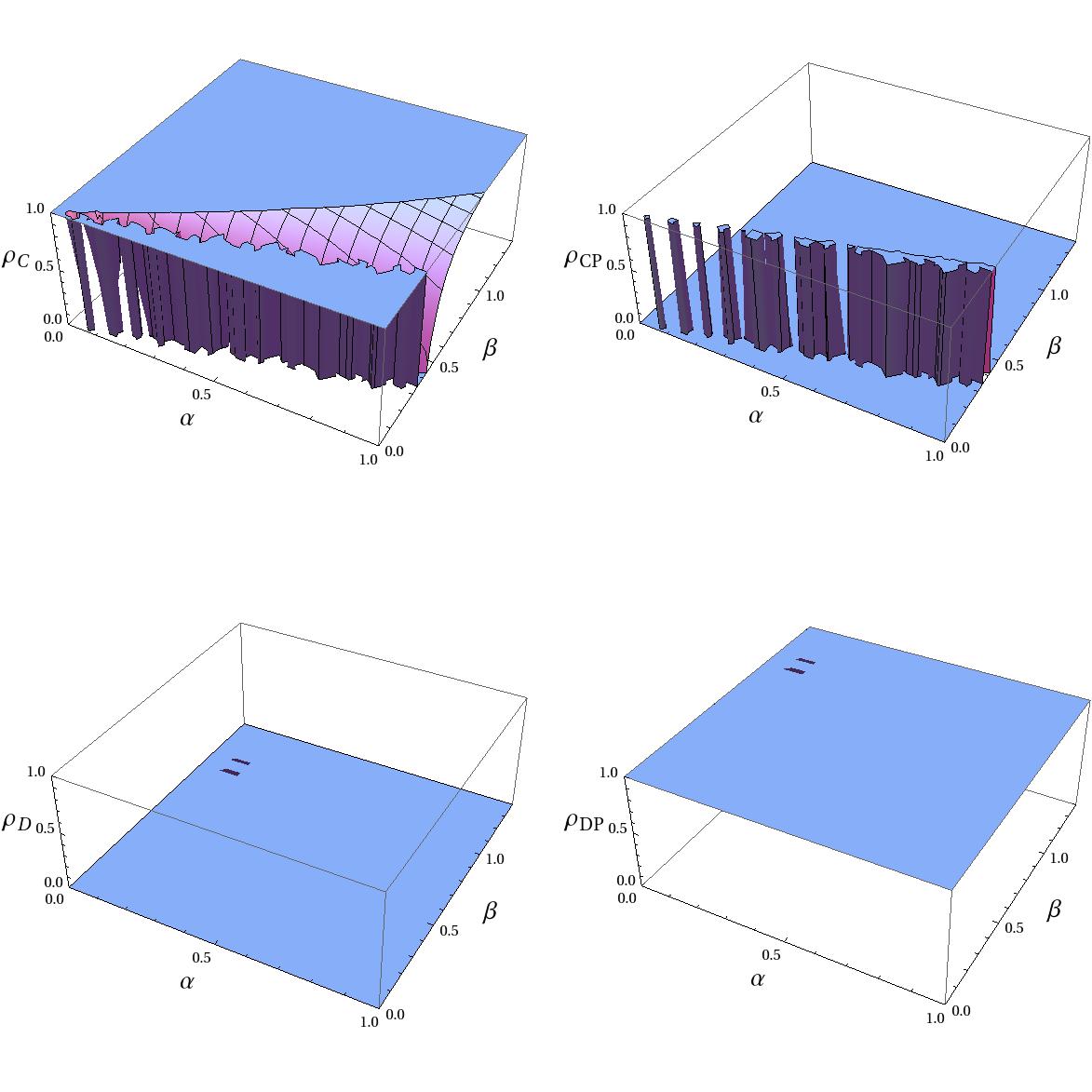}
\caption{The frequencies of four strategies based on spatial replicator equation, as function of $\alpha$ and $\beta$ for $b=1.4$ and $\gamma=0.8$. It can be observed that there are not parameter regions where the relationships $0<\rho_{C}<1$, $0<\rho_{D}<1$, $0<\rho_{CP}< 0$, and $0<\rho_{DP}< 0$ are satisfied at the same time. It also indicates that the coexistence of the four strategies is impossible.}
\label{four2}
\end{figure}
On basis of the spatial replicator equations considering the spatial structure of the population~(\cite{st1,st2}), the payoff for the four strategies are
$P_{s_{i}}  =  \sum\limits^{4}_{j=1}\rho_{s_{j}}c_{ij}$. %$\rho_{s}$ denotes the frequency of corresponding strategy $s$ ($s\in \{C, D, CP, DP\}$).
We assume that the four strategies can coexist with each other, i.e., the relationship $P_{C}=P_{CP}=P_{D}=P_{DP}$ is satisfied. With the normalization condition $\sum\limits^{4}_{i=1}\rho_{s_{i}}=1$, we can plot the frequencies of all four strategies in Fig.~\ref{four1} with $b=1.6$ and $\gamma=0.2$, in Fig.~\ref{four2} with $b=1.4$ and $\gamma=0.8$. It suggests that the four-strategy phase is impossible in the population.

By means of the same method, we can determine whether a three-strategy phase (i.e., three strategies can finally coexist stably together) can emerge. Accordingly, there are four different cases as follows.

(i) For C+CP+D phase:\\
$\bordermatrix{%
& C & CP & D \cr
C   & 0 & 0 & \frac{1}{2}(1-b)  \cr
CP  & 0 & 0 & \frac{1}{2}(1-b)+\frac{1}{2}(\beta-3\alpha)  \cr
D  & \frac{3}{2}(b-1) & \frac{3}{2}(b-1)+\frac{1}{2}(\alpha-3\beta)  & 0 \cr
}$ is the new payoff matrix.
Combing with the closure condition $\rho_{C}+\rho_{CP}+\rho_{D}=1$ and coexisting condition $P_{C}=P_{CP}=P_{D}$, we obtain fractions of the three strategies $\rho_{C}=1+\frac{3(b-1)}{\alpha-3\beta}$, $\rho_{CP}=\frac{3(1-b)}{\alpha-3\beta}$, and $\rho_{D}=0$. Theoretically, it suggests that the three-strategy phase is impossible in the system.

(ii) For C+CP+DP phase: \\
$\bordermatrix{%
& C & CP & DP \cr
C   & 0 & 0 &   \frac{1}{2}(1-b)+\frac{3}{2}(\alpha+\beta)\gamma   \cr
CP  & 0 & 0 & \frac{1}{2}(1-b)+2\beta \gamma \cr
DP  & \frac{3}{2}(b-1)-\frac{1}{2}(\alpha+\beta)\gamma  & \frac{3}{2}(b-1)-2\beta\gamma & 0\cr
}$ is the corresponding new payoff matrix.
Combing with the closure condition $\rho_{C}+\rho_{CP}+\rho_{DP}=1$ and coexisting condition $P_{C}=P_{CP}=P_{DP}$, we obtain fractions of the three strategies $\rho_{C}=1-\frac{3(b-1)-2\beta \gamma}{(\alpha-3\beta)\gamma}$, $\rho_{CP}=\frac{b-1}{(\alpha-3\beta)\gamma}$, and $\rho_{DP}=0$. Theoretically, it suggests that the three-strategy phase is impossible in the system. 

\begin{figure}[htbp]
\centering
\includegraphics[width=\textwidth,height=7cm]{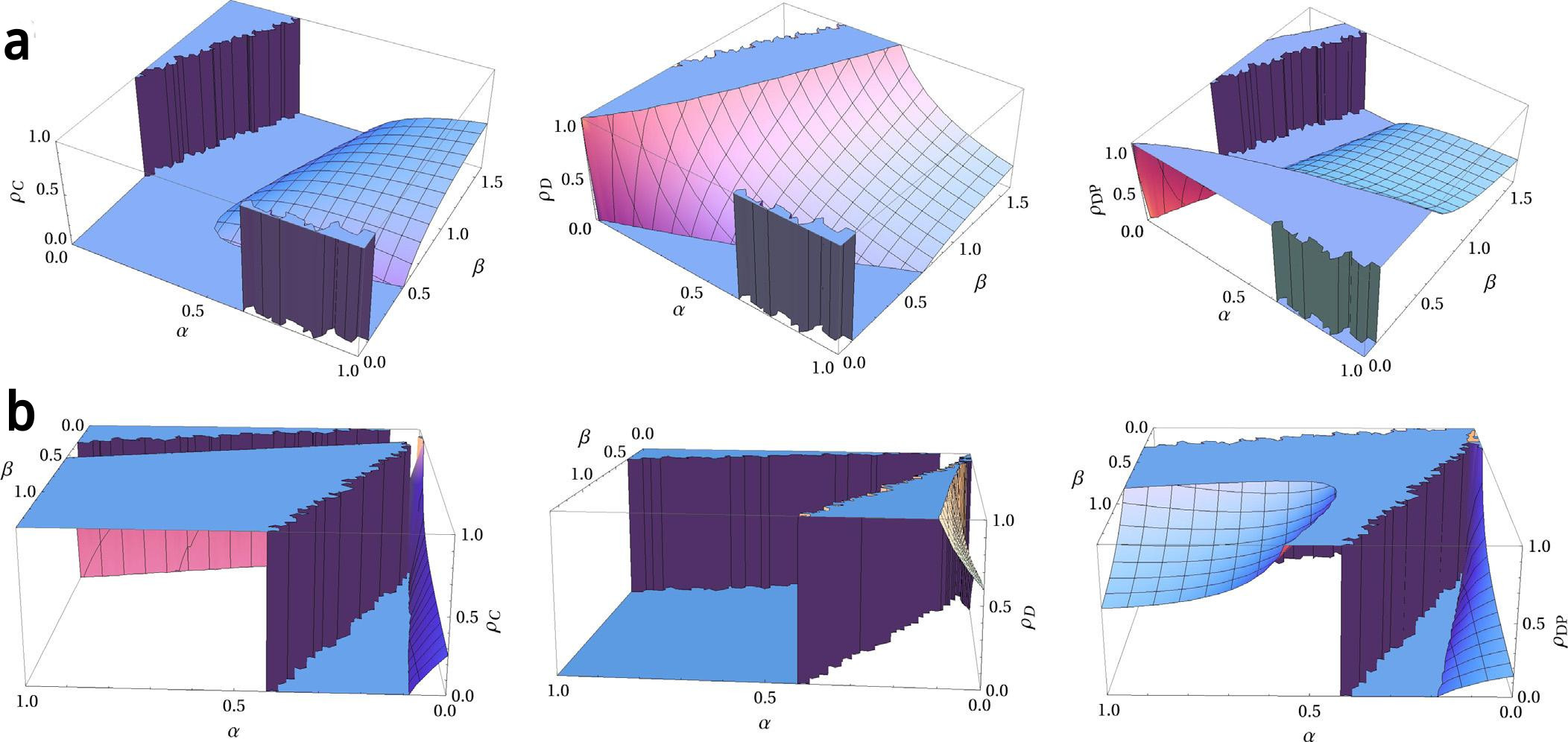}
\caption{The frequencies of the three strategies C, D, and DP as function of $\alpha$ and $\beta$. The parameters are taken as a. $b=1.6$ and $\gamma=0.2$; b. $b=1.4$ and $\gamma=0.8$. In either parameter condition, it can be found that there exists a parameter region where the relationships $0<\rho_{C}<1$, $0<\rho_{D}<1$, and $0<\rho_{DP}< 1$ are satisfied at the same time. It indicates that there are the stable coexistence of the three strategies under the two parameter conditions.}
\label{cddp}
\end{figure}
(iii) For C+D+DP phase:\\
$\bordermatrix{%
& C  & D & DP \cr
C   & 0 & \frac{1}{2}(1-b)  &   \frac{1}{2}(1-b)+\frac{3}{2}(\alpha+\beta)\gamma   \cr
D  & \frac{3}{2}(b-1)  & 0  & \frac{3}{2}(\alpha+\beta)\gamma +\frac{1}{2}(\alpha-3\beta) \cr
DP  & \frac{3}{2}(b-1)-\frac{1}{2}(\alpha+\beta)\gamma  & \frac{1}{2}(\beta-3\alpha)-\frac{1}{2}(\alpha+\beta)\gamma & 0\cr
}$ is the corresponding new payoff matrix.
Combing with the closure condition $\rho_{C}+\rho_{D}+\rho_{DP}=1$ and coexisting condition $P_{C}=P_{D}=P_{DP}$, we plot the profiles of $\rho_{C}$, $\rho_{D}$, and $\rho_{DP}$ in Fig.~\ref{cddp}  for $b=1.6$, $\gamma=0.2$ and $b=1.4$ and $\gamma=0.8$ as function of $\alpha$ and $\beta$. Theoretically, it suggests that the three-strategy phase would emerge conditionally in the system. 

\begin{figure}[htbp]
\centering
\includegraphics[width=\textwidth,,height=7cm]{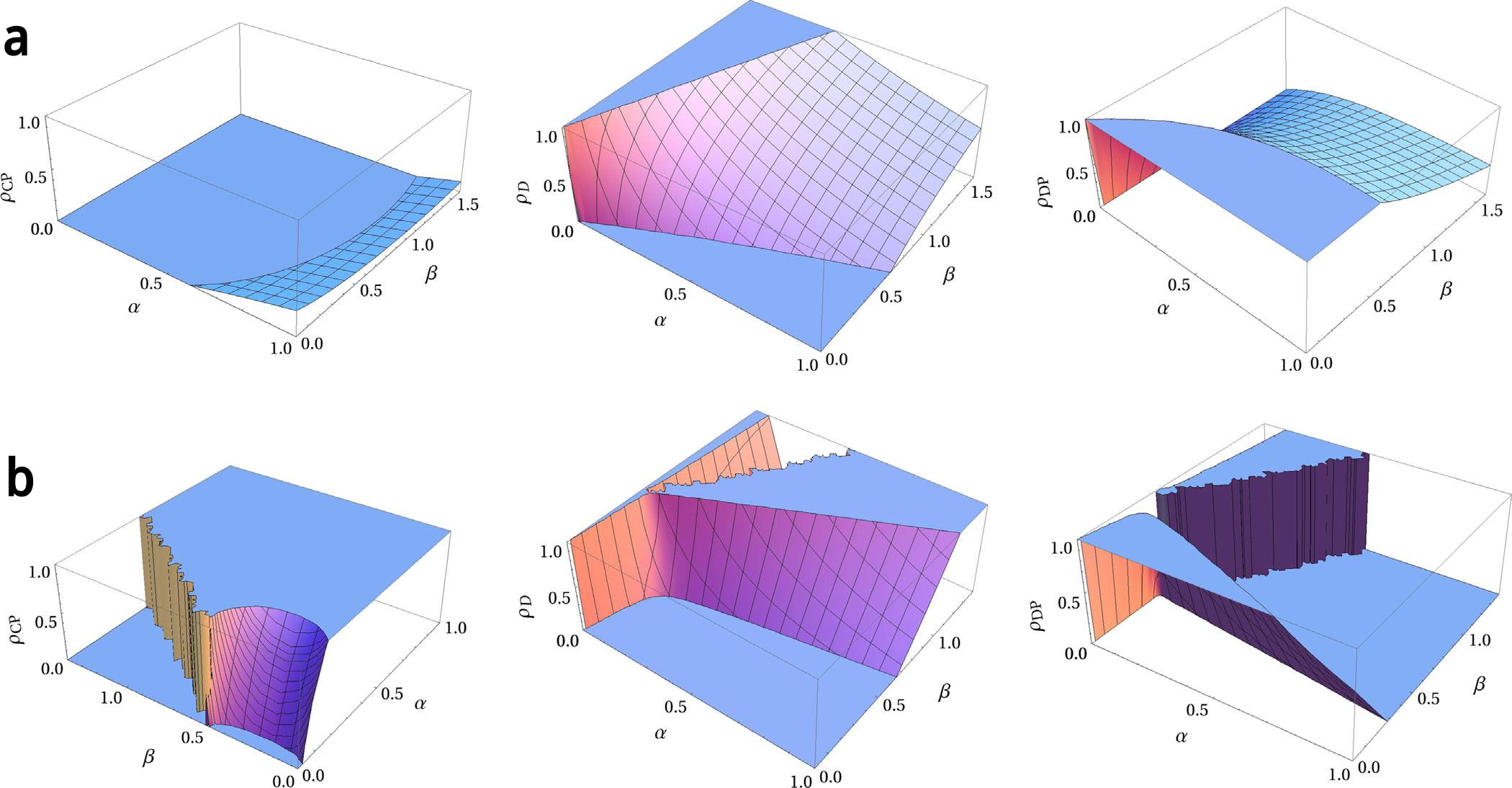}[h]
\caption{The frequencies of the three strategies CP, D, and DP as function of $\alpha$ and $\beta$. The parameters are taken as a. $b=1.6$ and $\gamma=0.2$; b. $b=1.4$ and $\gamma=0.8$. It can be found that there exists a parameter region where the relationships $0<\rho_{CP}<1$, $0<\rho_{D}<1$, and $0<\rho_{DP}< 1$ are satisfied at the same time. There is the stable coexistence of the three strategies for $b=1.6$ and $\gamma=0.2$. In b., it can be observed that there are not such parameter regions where the relationships $0<\rho_{CP}<1$, $0<\rho_{D}<1$, and $0<\rho_{DP}< 1$ are satisfied at the same time. The stable coexistence of the three strategies is impossible for $b=1.4$ and $\gamma=0.8$.}
\label{cpddp}
\end{figure}
(iv) For CP+D+DP phase:\\
$\bordermatrix{%
& CP & D & DP \cr
CP & 0 & \frac{1}{2}(1-b)+\frac{1}{2}(\beta-3\alpha)   & \frac{1}{2}(1-b)+2\beta \gamma \cr
D  & \frac{3}{2}(b-1)+\frac{1}{2}(\alpha-3\beta)  & 0  & \frac{3}{2}(\alpha+\beta)\gamma +\frac{1}{2}(\alpha-3\beta) \cr
DP & \frac{3}{2}(b-1)-2\beta\gamma  & \frac{1}{2}(\beta-3\alpha)-\frac{1}{2}(\alpha+\beta)\gamma & 0\cr
}$ is the corresponding new payoff matrix.
Combing with the closure condition $\rho_{CP}+\rho_{D}+\rho_{DP}=1$ and co-exist condition $P_{CP}=P_{D}=P_{DP}$, we plot the profiles of $\rho_{CP}$, $\rho_{D}$, and $\rho_{DP}$ in Fig.~\ref{cpddp}  for $b=1.6$, $\gamma=0.2$ and $b=1.4$ and $\gamma=0.8$ as function of $\alpha$ and $\beta$. Theoretically, it suggests that the three-strategy phase would emerge conditionally in the system for $b=1.6$ and $\gamma=0.2$. 

\section*{Appendix C: The analysis of interaction webs for $b=1.4$ and $\gamma=0.8$}
\label{apc}
\begin{figure}[htbp]
\centering
\includegraphics[width=\textwidth]{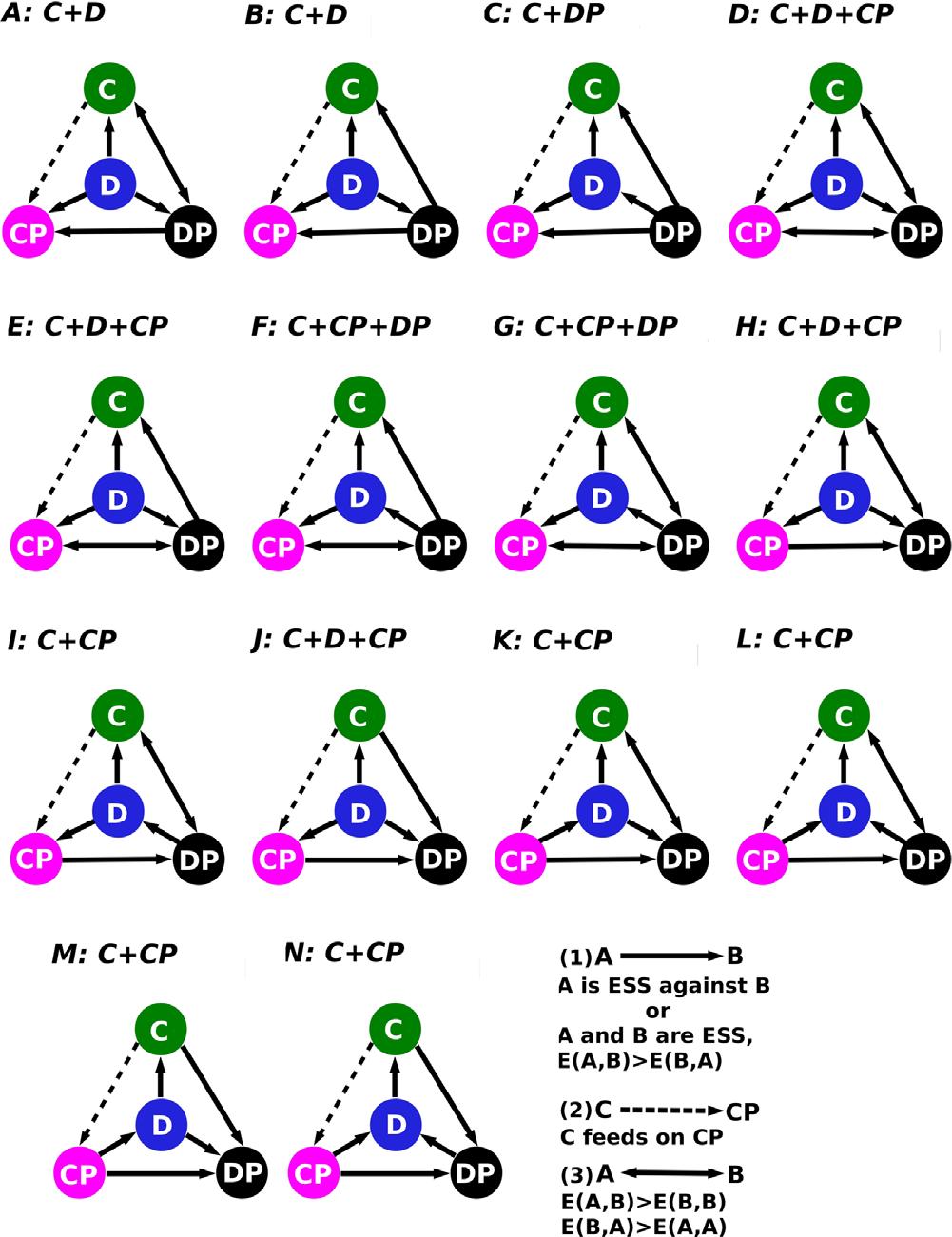}[hp]
\caption{The interaction graphs for the stable existence shown in Fig~\ref{148}, including the induced predictions. The inducing rules are the same as that illustrated at the caption of Fig.~\ref{web1}. The predictions just does not correspond with the real coexistence in regions B, C, E, near the junction of the line $\beta=\frac{1+\gamma}{1-\gamma}\alpha$, the line $\beta=-\alpha+\frac{b-1}{3\gamma}$, and the line $\beta=\frac{3(b-1)}{4\gamma}$.}
\label{web2}
\end{figure}

 \section*{References}
\bibliographystyle{elsarticle-harv}
\bibliography{selfish_reference}

\begin{thebibliography}{58}
\expandafter\ifx\csname natexlab\endcsname\relax\def\natexlab#1{#1}\fi
\expandafter\ifx\csname url\endcsname\relax
  \def\url#1{\texttt{#1}}\fi
\expandafter\ifx\csname urlprefix\endcsname\relax\def\urlprefix{URL }\fi

\bibitem[{Adami et~al.(2012)Adami, Schossau, and Hintze}]{original}
Adami, C., Schossau, J., Hintze, A., 2012. Evolution and stability of altruist
  strategies in microbial games. Phys. Rev. E 85~(1), 011914.

\bibitem[{Alexander(1987)}]{reputation1}
Alexander, R.~D., 1987. The Biology of Moral Systems. Transaction Books.

\bibitem[{Antal et~al.(2009)Antal, Nowak, and Traulsen}]{mutation}
Antal, T., Nowak, M.~A., Traulsen, A., 2009. Strategy abundance in 2$\times$ 2
  games for arbitrary mutation rates. J. Theor. Biol. 257~(2), 340--344.

\bibitem[{Axelrod(1984)}]{direct}
Axelrod, R.~M., 1984. The evolution of cooperation. Basic books, New York.

\bibitem[{Belot et~al.(2012)Belot, Bhaskar, and Van De~Ven}]{lie4}
Belot, M., Bhaskar, V., Van De~Ven, J., 2012. Can observers predict
  trustworthiness? Review of Economics and Statistics 94~(1), 246--259.

\bibitem[{Borjas(1989)}]{migrationcost1}
Borjas, G.~J., 1989. Economic theory and international migration. The
  International migration review 23~(3), 457--485.

\bibitem[{Boyd and Richerson(1992)}]{largep1}
Boyd, R., Richerson, P.~J., 1992. Punishment allows the evolution of
  cooperation (or anything else) in sizable groups. Ethology and Sociobiology
  13~(3), 171--195.

\bibitem[{Brandt et~al.(2003)Brandt, Hauert, and Sigmund}]{prsb}
Brandt, H., Hauert, C., Sigmund, K., 2003. Punishment and reputation in spatial
  public goods games. Proceedings of the Royal Society of London. Series B:
  Biological Sciences 270~(1519), 1099--1104.

\bibitem[{Buesser et~al.(2013)Buesser, Tomassini, and
  Antonioni}]{migrationcost4}
Buesser, P., Tomassini, M., Antonioni, A., 2013. Opportunistic migration in
  spatial evolutionary games. Phys. Rev. E 88~(4), 042806.

\bibitem[{Dercole et~al.(2013)Dercole, De~Carli, Della~Rossa, and
  Papadopoulos}]{mix2}
Dercole, F., De~Carli, M., Della~Rossa, F., Papadopoulos, A.~V., 2013.
  Overpunishing is not necessary to fix cooperation in voluntary public goods
  games. J. Theor. Biol. 326~(7), 70--81.

\bibitem[{Dobramysl and T{\"a}uber(2013)}]{prey}
Dobramysl, U., T{\"a}uber, U.~C., 2013. Environmental versus demographic
  variability in two-species predator-prey models. Phys. Rev. Lett. 110~(4),
  048105.

\bibitem[{Drinkwater et~al.(2003)Drinkwater, Levine, Lotti, Pearlman, and
  Welt-Wirtschafts-Archiv}]{migrationcost2}
Drinkwater, S., Levine, P., Lotti, E., Pearlman, J., Welt-Wirtschafts-Archiv,
  H., 2003. The economic impact of migration: A survey. Hamburgisches
  Welt-Wirtschafts-Archiv (HWWA).

\bibitem[{Eldakar et~al.(2007)Eldakar, Farrell, and Wilson}]{selfishpunish2}
Eldakar, O.~T., Farrell, D.~L., Wilson, D.~S., 2007. Selfish punishment:
  altruism can be maintained by competition among cheaters. J. Theor. Biol.
  249~(2), 198--205.

\bibitem[{Falconer et~al.(2011)Falconer, Czarny, and Brown}]{micro}
Falconer, S.~B., Czarny, T.~L., Brown, E.~D., 2011. Antibiotics as probes of
  biological complexity. Nat. Chem. Biol. 7~(7), 415--423.

\bibitem[{Fehr and G{\"a}chter(2002)}]{Alpunishment}
Fehr, E., G{\"a}chter, S., 2002. Altruistic punishment in humans. Nature
  415~(6868), 137--140.

\bibitem[{Fehr and Rockenbach(2003)}]{sdilemmas1}
Fehr, E., Rockenbach, B., 2003. Detrimental effects of sanctions on human
  altruism. Nature 422~(6928), 137--140.

\bibitem[{Foster et~al.(2006)Foster, Wenseleers, Ratnieks, and Queller}]{kin2}
Foster, K.~R., Wenseleers, T., Ratnieks, F.~L., Queller, D.~C., 2006. There is
  nothing wrong with inclusive fitness. Trends. Ecol. Evol. 21~(11), 599--600.

\bibitem[{Fowler(2005)}]{sdilemmas2}
Fowler, J.~H., 2005. Altruistic punishment and the origin of cooperation. Proc.
  Natl. Acad. Sci. 102~(19), 7047--7049.

\bibitem[{Fowler and Christakis(2010)}]{Alpunishment6}
Fowler, J.~H., Christakis, N.~A., 2010. Cooperative behavior cascades in human
  social networks. Proc. Natl. Acad. Sci. 107~(12), 5334--5338.

\bibitem[{Fu et~al.(2010)Fu, Nowak, and Hauert}]{pair1}
Fu, F., Nowak, M.~A., Hauert, C., 2010. Invasion and expansion of cooperators
  in lattice populations: Prisoner's dilemma vs. snowdrift games. J. Theor.
  Biol. 266~(3), 358--366.

\bibitem[{G{\'o}mez-Garde{\~n}es et~al.(2007)G{\'o}mez-Garde{\~n}es, Campillo,
  Flor{\'\i}a, and Moreno}]{matrix4}
G{\'o}mez-Garde{\~n}es, J., Campillo, M., Flor{\'\i}a, L., Moreno, Y., 2007.
  Dynamical organization of cooperation in complex topologies. Phys. Rev. Lett.
  98~(10), 108103.

\bibitem[{Haley and Fessler(2005)}]{reputation2}
Haley, K.~J., Fessler, D.~M., 2005. Nobody's watching?: Subtle cues affect
  generosity in an anonymous economic game. Evol. Hum. Behav. 26~(3), 245--256.

\bibitem[{Hamilton(1971)}]{kin1}
Hamilton, W.~D., 1971. Geometry for the selfish herd. J. Theor. Biol. 31~(2),
  295--311.

\bibitem[{Hauert et~al.(2007)Hauert, Traulsen, Brandt, Nowak, and
  Sigmund}]{nowak}
Hauert, C., Traulsen, A., Brandt, H., Nowak, M.~A., Sigmund, K., 2007. Via
  freedom to coercion: the emergence of costly punishment. Science 316~(5833),
  1905--1907.

\bibitem[{Helbing et~al.(2010{\natexlab{a}})Helbing, Szolnoki, Perc, and
  Szab{\'o}}]{cost1}
Helbing, D., Szolnoki, A., Perc, M., Szab{\'o}, G., 2010{\natexlab{a}}.
  Defector-accelerated cooperativeness and punishment in public goods games
  with mutations. Phys. Rev. E 81~(5), 057104.

\bibitem[{Helbing et~al.(2010{\natexlab{b}})Helbing, Szolnoki, Perc, and
  Szab{\'o}}]{spatial2}
Helbing, D., Szolnoki, A., Perc, M., Szab{\'o}, G., 2010{\natexlab{b}}.
  Evolutionary establishment of moral and double moral standards through
  spatial interactions. PloS Comput. Biol. 6~(4), e1000758.

\bibitem[{Helbing et~al.(2010{\natexlab{c}})Helbing, Szolnoki, Perc, and
  Szab{\'o}}]{cost2}
Helbing, D., Szolnoki, A., Perc, M., Szab{\'o}, G., 2010{\natexlab{c}}. Punish,
  but not too hard: how costly punishment spreads in the spatial public goods
  game. New J. Phys. 12~(8), 083005.

\bibitem[{Helbing and Yu(2009)}]{migration1}
Helbing, D., Yu, W., 2009. The outbreak of cooperation among success-driven
  individuals under noisy conditions. Proc. Natl. Acad. Sci. 106~(10),
  3680--3685.

\bibitem[{Killingback et~al.(2006)Killingback, Bieri, and Flatt}]{net3}
Killingback, T., Bieri, J., Flatt, T., 2006. Evolution in group-structured
  populations can resolve the tragedy of the commons. Proc. Natl. Acad. Sci.
  273~(1593), 1477--1481.

\bibitem[{Knebel et~al.(2013)Knebel, Kr{\"u}ger, Weber, and Frey}]{lv}
Knebel, J., Kr{\"u}ger, T., Weber, M.~F., Frey, E., 2013. Coexistence and
  survival in conservative lotka-volterra networks. Phys. Rev. Lett. 110~(16),
  168106.

\bibitem[{Lieberman et~al.(2005)Lieberman, Hauert, and Nowak}]{nowaknet}
Lieberman, E., Hauert, C., Nowak, M.~A., 2005. Evolutionary dynamics on graphs.
  Nature 433~(7023), 312--316.

\bibitem[{Martin(2012)}]{migrationcost3}
Martin, P., 2012. The economic analysis of international migration and
  migration policy: Toward a research agenda. Discussion Paper prepared for 22
  May 2012 Migration Impacts Symposium held at COMPAS, University of Oxford.

\bibitem[{Mathew and Boyd(2011)}]{history}
Mathew, S., Boyd, R., 2011. Punishment sustains large-scale cooperation in
  prestate warfare. Proc. Natl. Acad. Sci. 108~(28), 11375--11380.

\bibitem[{Nakamaru and Iwasa(2006)}]{spatial1}
Nakamaru, M., Iwasa, Y., 2006. The coevolution of altruism and punishment: role
  of the selfish punisher. J. Theor. Biol. 240~(3), 475--488.

\bibitem[{Nowak(2006)}]{allfor}
Nowak, M.~A., 2006. Five rules for the evolution of cooperation. redScience
  314~(5805), 1560--1563.

\bibitem[{Nowak and May(1992)}]{norm1}
Nowak, M.~A., May, R.~M., 1992. Evolutionary games and spatial chaos. Nature
  359~(6398), 826--829.

\bibitem[{Nowak and Sigmund(1998{\natexlab{a}})}]{indirect2}
Nowak, M.~A., Sigmund, K., 1998{\natexlab{a}}. The dynamics of indirect
  reciprocity. J. Theor. Biol. 194~(4), 561--574.

\bibitem[{Nowak and Sigmund(1998{\natexlab{b}})}]{indirect1}
Nowak, M.~A., Sigmund, K., 1998{\natexlab{b}}. Evolution of indirect
  reciprocity by image scoring. Nature 393~(6685), 573--577.

\bibitem[{Nowak et~al.(2010)Nowak, Tarnita, and Antal}]{net2}
Nowak, M.~A., Tarnita, C.~E., Antal, T., 2010. Evolutionary dynamics in
  structured populations. Philos. T. R. Soc. B. 365~(1537), 19--30.

\bibitem[{Ohtsuki et~al.(2006)Ohtsuki, Hauert, Lieberman, and
  Nowak}]{cooperation2}
Ohtsuki, H., Hauert, C., Lieberman, E., Nowak, M.~A., 2006. A simple rule for
  the evolution of cooperation on graphs and social networks. Nature
  441~(7092), 502--505.

\bibitem[{Ohtsuki and Nowak(2006)}]{st1}
Ohtsuki, H., Nowak, M.~A., 2006. The replicator equation on graphs. J. Theor.
  Biol. 243~(1), 86--97.

\bibitem[{Ohtsuki and Nowak(2008)}]{st2}
Ohtsuki, H., Nowak, M.~A., 2008. Evolutionary stability on graphs. J. Theor.
  Biol. 251~(4), 698--707.

\bibitem[{Panchanathan and Boyd(2004)}]{largep2}
Panchanathan, K., Boyd, R., 2004. Indirect reciprocity can stabilize
  cooperation without the second-order free rider problem. Nature 432~(7016),
  499--502.

\bibitem[{Parkhe(1993)}]{alliance2}
Parkhe, A., 1993. Strategic alliance structuring: A game theoretic and
  transaction cost examination of interfirm cooperation. Academy of management
  journal 36~(4), 794--829.

\bibitem[{Perc(2012)}]{nonstable}
Perc, M., 2012. Sustainable institutionalized punishment requires elimination
  of second-order free-riders. Sci. Rep. 2, 344.

\bibitem[{Santos and Pacheco(2005)}]{matrix3}
Santos, F.~C., Pacheco, J.~M., 2005. Scale-free networks provide a unifying
  framework for the emergence of cooperation. Phys. Rev. Lett. 95~(9), 098104.

\bibitem[{Santos et~al.(2008)Santos, Santos, and Pacheco}]{cooperation3}
Santos, F.~C., Santos, M.~D., Pacheco, J.~M., 2008. Social diversity promotes
  the emergence of cooperation in public goods games. Nature 454~(7201),
  213--216.

\bibitem[{Serra-Garcia et~al.(2013)Serra-Garcia, Van~Damme, and Potters}]{lie2}
Serra-Garcia, M., Van~Damme, E., Potters, J., 2013. Lying about what you know
  or about what you do? Journal of the European Economic Association 11~(5),
  1204--1229.

\bibitem[{Sheremeta and Shields(2013)}]{lie3}
Sheremeta, R.~M., Shields, T.~W., 2013. Do liars believe? beliefs and
  other-regarding preferences in sender--receiver games. Journal of Economic
  Behavior \& Organization 94, 268--277.

\bibitem[{Sigmund et~al.(2010)Sigmund, De~Silva, Traulsen, and
  Hauert}]{sdilemmas3}
Sigmund, K., De~Silva, H., Traulsen, A., Hauert, C., 2010. Social learning
  promotes institutions for governing the commons. Nature 466~(7308), 861--863.

\bibitem[{Szab{\'o} and T{\H{o}}ke(1998)}]{imitate}
Szab{\'o}, G., T{\H{o}}ke, C., 1998. Evolutionary prisoner’s dilemma game on
  a square lattice. Phys. Rev. E 58~(1), 69.

\bibitem[{Szab{\'o} et~al.(2005)Szab{\'o}, Vukov, and Szolnoki}]{monotonic1}
Szab{\'o}, G., Vukov, J., Szolnoki, A., 2005. Phase diagrams for an
  evolutionary prisoner’s dilemma game on two-dimensional lattices. Phys.
  Rev. E 72~(4), 047107.

\bibitem[{Szolnoki et~al.(2011)Szolnoki, Szab{\'o}, and Perc}]{cost3}
Szolnoki, A., Szab{\'o}, G., Perc, M., 2011. Phase diagrams for the spatial
  public goods game with pool punishment. Phys. Rev. E 83~(3), 036101.

\bibitem[{Tarnita et~al.(2009)Tarnita, Antal, Ohtsuki, and Nowak}]{monotonic2}
Tarnita, C.~E., Antal, T., Ohtsuki, H., Nowak, M.~A., 2009. Evolutionary
  dynamics in set structured populations. Proc. Natl. Acad. Sci. 106~(21),
  8601--8604.

\bibitem[{Todeva and Knoke(2005)}]{alliance1}
Todeva, E., Knoke, D., 2005. Strategic alliances and models of collaboration.
  Management Decision 43~(1), 123--148.

\bibitem[{Wilson and O’Gorman(2006)}]{selfishpunish}
Wilson, D.~S., O’Gorman, R., 2006. Emotions and actions associated with
  altruistic helping and punishment. Evolutionary Psychology 4, 274--286.

\bibitem[{Wolff(2012)}]{mix0}
Wolff, I., 2012. Retaliation and the role for punishment in the evolution of
  cooperation. J. Theor. Biol. 315, 128--138.

\bibitem[{Yang et~al.(2010)Yang, Wu, and Wang}]{migration2}
Yang, H.-X., Wu, Z.-X., Wang, B.-H., 2010. Role of aspiration-induced migration
  in cooperation. Phys. Rev. E 81~(6), 065101.

\end{thebibliography}




%% \bibitem must have one of the following forms:
%%   \bibitem[Jones et al.(1990)]{key}...
%%   \bibitem[Jones et al.(1990)Jones, Baker, and Williams]{key}...
%%   \bibitem[Jones et al., 1990]{key}...
%%   \bibitem[\protect\citeauthoryear{Jones, Baker, and Williams}{Jones
%%       et al.}{1990}]{key}...
%%   \bibitem[\protect\citeauthoryear{Jones et al.}{1990}]{key}...
%%   \bibitem[\protect\astroncite{Jones et al.}{1990}]{key}...
%%   \bibitem[\protect\citename{Jones et al., }1990]{key}...
%%   \harvarditem[Jones et al.]{Jones, Baker, and Williams}{1990}{key}...
%%


% \end{thebibliography}

\end{document}